\documentstyle[preprint,psfig,aps]{revtex}
\begin{document}
\def\e{\mbox{e}}
\def\sgn{{\rm sgn}}
\def\gsim{\;
\raise0.3ex\hbox{$>$\kern-0.75em\raise-1.1ex\hbox{$\sim$}}\;
}
\def\lsim{\;
\raise0.3ex\hbox{$<$\kern-0.75em\raise-1.1ex\hbox{$\sim$}}\;
}
\def\MeV{\rm MeV}
\def\eV{\rm eV}
\draft

\title{ Neutrino conversions in random magnetic fields
and $\tilde{\nu}_e$ from the Sun}

\author{ A.A. Bykov~$^a$~\footnote{E-mail: bikov@math380b.phys.msu.su},
 V.Yu. Popov~$^a$~\footnote{E-mail: popov@math380b.phys.msu.su},
 A.I. Rez~$^b$~\footnote{E-mail: rez@izmiran.rssi.ru},\\
 V.B. Semikoz~$^c$~\footnote{E-mail: semikoz@flamenco.ific.uv.es~.
On leave from IZMIRAN, Troitsk, Moscow region, 142092, Russia} and
 D.D. Sokoloff~$^a$~\footnote{E-mail: sokoloff@lem.srcc.msu.su}}

\address{$^a$ Department of Physics, Moscow State University,
119899, Moscow, Russia}
\address{$^b$ The Institute of the Terrestrial Magnetism, the Ionosphere and
Radio Wave\\ Propagation of the Russian Academy of Sciences,
             IZMIRAN,\\Troitsk, Moscow region, 142092, Russia}
\address{$^c$ Instituto de F\'{\i}sica Corpuscular - C.S.I.C.,
Departament de F\'{\i}sica Te\`orica\\ Universitat de Val\`encia,
46100 Burjassot, Val\`encia, SPAIN }
\tighten
\maketitle
\begin{abstract}
The magnetic field in the convective zone of the Sun has a random
small-scale component with the r.m.s.  value substantially exceeding
the strength of a regular large-scale field.
For two Majorana neutrino flavors $\times$ two helicities in the
presence of a neutrino transition magnetic moment and nonzero neutrino
mixing we analyze the displacement of the allowed ($\Delta
m^2$-$\sin^2\theta$)- parameter region reconciled for the SuperKamiokande(SK)
and radiochemical (GALLEX, SAGE, Homestake) experiments
in dependence on the r.m.s. magnetic field value $b$, or more precisely,
on a value $\mu b$ assuming the transition magnetic moment
$\mu = 10^{-11}\mu_B$. In contrast to RSFP in regular magnetic fields
we find an effective production of electron
antineutrinos in the Sun even for small neutrino mixing
through the cascade conversions $\nu_{eL}\to \nu_{\mu L}\to
\tilde{\nu}_{eR}$, $\nu_{eL}\to \bar{\nu}_{\mu R}\to \bar{\nu}_{eR}$
in a random magnetic field that would be a signature of the Majorana
nature of neutrino if $\tilde{\nu}_{eR}$ will be registered.
Basing on the present SK bound on electron antineutrinos we have also
found an excluded area in the same $\Delta m^2$, $\sin^22\theta$ -plane
and revealed a strong sensitivity to the random magnetic field
correlation length $L_0$.
\end{abstract}


\section{Introduction}

Recent results of the SuperKamiokande (SK) experiment\cite{SK} have
already confirmed the solar neutrino deficit at the level less
$R_{SK}/R_{SSM}\lsim 0.4$.

Moreover, day/night (D/N) and
season neutrino flux variations were analized in this experiment and
in $\approx$ 4 years one expects to
reach
enough statistics in order to confirm or to refuse these periods
predicted in some theoretical models. D/N variations would be a signature
of the MSW solution to the Solar Neutrino Problem (SNP). The signature of the
vacuum oscillations is a seasonal variation of neutrino flux, in addition
to geometrical seasonal variation $1/r^2(t)$. And signature of the resonant
spin-flavor precession (RSFP)-solution is the 11 year periodicity of
neutrino flux.

All three elementary particle physics solutions (vacuum, MSW and RSFP)
successfully describe the results of all four solar-neutrino
experiments, because the suppression factors of neutrino fluxes are
energy dependent and  helioseismic data confirm the Standard Solar Model
(SSM) with high precision at all radial distances of interest (see
recent review by Berezinsky \cite{Berezinsky}).

There exists, however, a problem with large regular solar magnetic fields,
essential for the RSFP scenario. It is commonly accepted that
magnetic fields measured at the surface of the Sun are weaker than
within interior of the convective zone where this field is supposed to be
generated. The mean field value over the solar disc is about of order
$1~ G$ and in the solar spots magnetic field strength reaches $ 1~ kG$.

Because sunspots are considered to be produced from magnetic tubes
transported to the solar surface due to the boyancy, this figure can be
considered as a reasonable order-of magnitude observational estimate for
the mean magnetic field strength in the region of magnetic field generation.
In the solar magnetohydrodynamics (see e.g. \cite{Parker}))
one can explain such fields in a self-consistent way if these fields are
generated by dynamo mechanism at the bottom of the convective zone (or,
more specific, in the overshoot layer). But its value
seems to be too low for effective neutrino conversions.

The mean magnetic field is however followed by a {\it small scale}, random
magnetic field. This random magnetic field is not directly traced by
sunspots or other tracers of solar activity.
This field propagates through convective zone and photosphere
drastically decreasing in the strength value with an increase of the
scale. According to the available understanding of solar dynamo, the
strength of the random magnetic field inside the convective zone is
larger than the mean field strength. A direct observational estimation
of the ratio between this strengthes is not available, however the ratio
of order 50 -- 100 does not seem impossible. At least, the ratio between
the mean magnetic field strength and the fluctuation at the solar surface
is estimated as 50 (see e.g. \cite{Stix}).

This is the main reason why we consider here an analogous to the RSFP
scenario, an aperiodic spin-flavour conversion (ASFC), based on the
presence of random magnetic fields in the solar convective zone.
It turns out that the ASFC is an additional probable way to describe
the solar neutrino deficit in different energy regions, especially if
current and future experiments will detect electron antineutrinos
from the Sun. The termin ``aperiodic'' simply reflects the exponential
behaviour of conversion probabilities in noisy media (cf. \cite{Nicolaidis}
, \cite{Enqvist}).

As well as for the RSFP mechanism\cite{Akhmedov} all arguments for
and against the ASFC mechanism with random magnetic fields remain the same
ones that have been recently summarized and commented by Akhmedov
(see \cite{Akhmedov1} and references therein).

But contrary to the case of regular magnetic fields we find out that
one of the signatures of random magnetic field in the Sun is the
prediction of a wider allowed region for neutrino mixing angle in the
presence of $\bar{\nu}_e$ from the Sun including the case of {\it
small mixing angles} (see below section IV).

Notice that if electron antineutrinos from the Sun were detected at
the Earth this would lead to conclusion that neutrinos are
Majorana particles and this is a very attractive point for applying  RSFP
in regular magnetic field or ASFC in random fields in the Sun.

The SK experiment provides a stringent bound on the presence of solar
electron antineutrinos at least for high energy region  \cite{Fiorentini}.
A phenomenological and numerical analysis of
$\nu_{eL}\to \tilde{\nu}_{eR}$-conversion in the solar twisting
magnetic fields \cite{APS} -  \cite{BL} shows that it is possible
to obtain a noticeable yield of  $\tilde{\nu}_{eR}~$ \cite{PSV}
consistent with the limit \cite{Fiorentini}. Twisting magnetic
fields themselves are, however, very specific and it is hard to
explain their existence and origin. Therefore, more realistic models of
magnetic field in the Sun are necessary and for a start we treat here
both SNP solution and $\tilde{\nu}_e$-production applying, as we hope,
a bit more realistic model of random magnetic fields in the convective
zone of the Sun.

The random magnetic field is considered to be
maximal somewhere at the bottom of convective zone and decaying to the
solar surface. To take into account a possibility, that the solar dynamo
action is possible also just below the bottom of the convective zone
(see \cite{Parker}),
we accept, rather arbitrary, that it is distributed at
the radial range $0.7 R_\odot \mbox{---} 1.0 R_\odot$, i.e. it has the same
thickness as the convective zone.
Of course, our model of the random magnetic field (see in
details below) is a crude simplification of the real situation,
however, the results  are presumably more or less robust.

For each realization of random
magnetic fields we find a solution of the Cauchi problem in the
form
of a set of wave functions $\nu_a(t) = \mid \nu_a(t)\mid \exp
(i\alpha_a(t))$
obeying the unitarity condition for the probabilities
$P_{aa}(t) = \nu_a^*(t)\nu_a(t)$,
\begin{equation}
P_{ee}(t) +
P_{\bar{e}\bar{e}}(t) + P_{\mu \mu}(t) + P_{\bar{\mu}\bar{\mu}}(t) =
1~,
\label{unitarity}
\end{equation}
where the subscript $a$ equals to $a=e$ for $\nu_{eL}$, $a= \bar{e}$
for $\bar{\nu}_{eR}$, $a = \mu$ for $\nu_{\mu L}$ and $a=\bar{\mu}$
for
$\bar{\nu}_{\mu R}$ correspondingly.

In section II the experimental observables are defined through the
neutrino conversion probabilities $P_{aa}(E)$ at the detectors
($r= R_{Earth}$) accounting for neutrino propagation  both in the Sun
and on the way from the Sun to the Earth.

In subsection II.A we give general formulas for the neutrino event
spectrum in real time experiments with measurement of recoil electron
energy in the $\nu e$-scattering.  In contrast to the well-known case
of regular fields, we need to find the probabilities $P_{aa}(r= t)$ as the
functions of a local position in the transversal plane too,
$P_{aa}(r)\equiv P_{aa}(r,\xi, \eta)$, because longitudinal profiles
of the random fields (along $r$) are generally different in different points
in the transversal plane even for an instant when parallel neutrino fluxes
directed to the Earth (called here "rays") cross the convective zone (see
Fig.\ \ref{fig01}).

As a result of different realizations of random fields along
different rays the probabilities $P_{aa}(r, \xi, \eta)$ are random
functions built on randomness of magnetic fields in all three
directions.

The same probabilities for left-handed electron neutrinos,
$P_{ee}(r,\xi , \eta)$, are used in subsection II.B to define
neutrino flux measured in radiochemical experiments
(Homestake and GALLEX + SAGE, in SNU units).

In main section III we give some physical arguments supporting the
random magnetic field model implemented into the master equation
Eq. (\ref{master}) (subsection III.B.1). After that we describe the
mathematical model of random magnetic fields (subsection III.B.2).

Then in subsection III.C we analyze asymptotic solution of the master
equation in the case of small neutrino mixing. We briefly discuss
there many possible analytic issues: in particular, magnetic field
correlations of finite radius (subsection III.C.2), linked
cluster expansion  and higher moments of survival
probability (subsection III.C.3).  We give also an interpretation of the
random magnetic field influence as a random walk over a circle
(subsection III.C.4).

In subsection III.D we describe the algorithm of our numerical approach.
The main goal is to calculate the mean arithmetic probabilities
as functions of  mixing parameters $\sin^22\theta$ , $\delta =
\Delta m^2/4E$,
\begin{equation}
\label{averprob}
\langle P_{aa}(\delta ,
\sin^22\theta)\rangle = \int \int d\xi d\eta P_{aa}(\delta,
\sin^22\theta, \xi, \eta )/\int \int d\xi d\eta \ ,
\end{equation}
under the assumption
$\Phi^{(0)}_i(\xi, \eta , E) = \Phi^{(0)}_i(\xi, \eta) \lambda_i (E) $,
where  $\lambda_i(E)$ is the
corresponding normalized differential flux,
$\int_0^{E_{max}(i)}\lambda_i(E)dE = 1$
and
$\Phi^{(0)}_i(\xi, \eta)$
is the integral flux of neutrinos of  kind "i" ($i =
pp~,Be~,pep~,B$)
assumed to be constant and uniform,
$\Phi^{(0)}_i(\xi, \eta) = \Phi^{(0)}_i = const $,
at a given distance $r = t$ from the center of the Sun \cite{Bahcall}.
This simplifies the averaging in the transversal plane since

\begin{equation}
\label{factor}
\int_{\xi^2 + \eta^2<
R_{core}^2}d\xi d\eta \Phi_i^{(0)}(\xi, \eta)P_{aa}(\xi, \eta)
\times [\int_{\xi^2 +
\eta^2< R_{core}^2}d\xi d\eta]^{-1} =
\Phi^{(0)}_i\times \langle P_{aa}\rangle \ .
\end{equation}

It is worth to note that the above averaging procedure merely reflects
the physical properties of neutrino detectors which response to the
incoherent sum of partial neutrino fluxes incoming from all visible
parts of the Sun. At the same time it is well known that the most
studied statistical characteristics in statistical physics and
especially in the theory of disordered media are those that are
additive functions of dimensions (cf. \cite{Lifshitz}). Their main
distinctive feature is that, being devided by corresponding volume,
they become certain in macroscopic limit, i.e. self-averaged. From
definition Eq. (\ref{averprob}) it follows that the
integral in the numerator in the r.h.s. is indeed an additive
function of the area of the convective zone layer (see also Fig.\ \ref{fig01})
and therefore for increasing area one can expect that
$\langle P_{aa}\rangle$ should be self-averaged. Our results below
confirm this property ( subsection III.C.3 and section IV).

In section IV we analyze how these probabilities depend on the r.m.s.  field
value $\sqrt{\langle b^2\rangle}$ and argue why electron antineutrinos
are not seen in the SK experiment reconciling available experimental
data for four solar neutrino experiments.

In final section V we discuss our results comparing two mechanisms, RSFP
and ASFC, for the same strength of regular and r.m.s. fields and
$\mu =10^{-11}\mu_B$.

\section{Experimental observable values}

If neutrinos have the transition magnetic moment, then
in real time experiments like $\nu$e-scattering in the SK case\cite{SK}
the recoil electron spectrum
depends on all four neutrino conversion probabilities, $P_{aa}$
(see subsection II.A),
while in the case of radiochemical experiments (Homestake, Gallex,
SAGE\cite{radiochem}) the number of events measured in SNU units depends
on charge current contribution with left-handed electron neutrinos only,
i.e. on the survival probability $P_{ee}$ that itself is a function of the
magnetic field parameter $\mu B_{\perp}$, as well as of the fundamental
mixing parameters $\Delta m^2$, $\sin^22\theta$ (subsection II.B).

\subsection{Spectrum and number of neutrino
events in $\nu e$-scattering experiments}

In a real time experiment one measures the integral spectrum (the number
of neutrino events per day)
\begin{equation}
N_{\nu} = \sum_{k_{min}}^{k_{max}}\int_{T_k}^{T_{k + 1}}dT
\frac{dN_{\nu}(T)}{dT}~,
\label{integral}
\end{equation}
where experimentalists devide the whole allowed recoil electron energy
interval $T\geq T_{th}$ into bins $\Delta W = T_{k + 1} - T_k$ with
$T_{k_{min}} = T_{th}$.
The energy spectrum of events, $dN_{\nu}/dT$, has the form

\widetext
\begin{eqnarray}
\frac{dN_{\nu}}{dT}&=&
N_e[\int
\int_{\xi^2 + \eta^2< R_{core}^2}d\xi d\eta]^{-1}\times
\int_{\xi^2 + \eta^2< R_{core}^2}d\xi d\eta\sum_i\Phi^{(0)}_i(\xi,
\eta)\times \nonumber\\
& &\int_{E_{min}(T)}^{E_{max}(i)}dE\varepsilon
(E)\lambda_i(E)\langle \frac{d\sigma^{(M)}_{weak}}{dT}(E, T||\xi,
\eta)\rangle~, \label{spectrum} \end{eqnarray}
where $N_e$ is the total
number of electrons in the fiducial volume of the detector;
$E_{min}(T) = (T/2)(1 + \sqrt{1 + 2m_e/T})$ is the minimal neutrino
energy obtained from the kinematical inequality $T\leq T_{max} =
2E^2/(2m_e + 2E)$ and $E_{max}(i)$ is given in \cite{Bahcall}. For
simplicity  the efficiency $\varepsilon (E)$
above the detector threshold $T_{th}$ is substituted by $\varepsilon
(E) = 1$.

Notice that we have assumed the same core radii for all neutrino
sources (kinds "i"), $R_{core} = R_{core}^{(i)}$.

The averaged
differential cross-section $\langle d\sigma^{(M)}_{weak}(E, T||\xi,
\eta)/dT\rangle $ in Eq.  (\ref{spectrum}),
\begin{eqnarray} \langle
\frac{d\sigma^{(M)}_{weak}}{dT}(E, T||\xi, \eta)\rangle = \Bigl
[\int_0^{T_{max}}\exp ( - (T - T^{'})^2/2\Delta
(T^{'})^2)dT^{'}\Bigr ]^{-1}\times \nonumber\\\times \Bigl
[\int_0^{T_{max}}\exp ( - (T - T^{'})^2/2\Delta
(T^{'})^2)dT^{'}\frac{d\sigma^{(M)}_{weak}(T^{'},
E||\xi,\eta)}{dT^{'}}\Bigr ]
\label{average}
\end{eqnarray}
is given by the energy resolution $\Delta T\equiv \Delta (T)$ in the
Gaussian distribution where $T^{'}$ is the {\it true} recoil electron energy,
$T$ is the measured energy, and by the $\nu e$ scattering
cross-section for four active Majorana neutrino components\cite{Sem97}

\begin{eqnarray}
\frac{d\sigma^{(M)}_{weak}(E, T^{'}||\xi, \eta)}{dT}
= \frac{2G_F^2m_e}{\pi} \{ P_{ee}(E, \xi, \eta)&& \Bigl [
g_{eL}^2 + g_R^2\Bigl (1 - \frac{T^{'}}{E}\Bigr )^2 -
\frac{m_eT^{'}}{E^2}g_{eL}g_R\Bigr ] + \nonumber\\  +
P_{\bar{e}\bar{e}}(E, \xi, \eta)&& \Bigl [g_R^2 + g_{eL}^2 \Bigl (1 -
\frac{T^{'}}{E}\Bigr )^2 - \frac{m_eT^{'}}{E^2}g_{eL}g_R\Bigr ]
+ \nonumber\\  + P_{\mu \mu}(E, \xi, \eta)&&\Bigl [ g_{\mu L}^2 +
 g_R^2\Bigl (1 - \frac{T^{'}}{E}\Bigr )^2 -
\frac{m_eT^{'}}{E^2}g_{\mu
 L}g_R\Bigr ] + \nonumber\\  +
P_{\bar{\mu}\bar{\mu}}(E, \xi, \eta)&& \Bigl [g_R^2 + g_{\mu L}^2 \Bigl
(1 -
 \frac{T^{'}}{E}\Bigr )^2 - \frac{m_eT^{'}}{E^2}g_{\mu
L}g_R\Bigr ] \}~.
\label{weak}
\end{eqnarray}
\narrowtext

Here
$g_{eL} = \xi + 0.5$, $g_{\mu L} = \xi - 0.5$, $g_R = \xi
= \sin^2 2 \theta_W  \simeq 0.23$ are
the coupling constants in the Standard Model(SM).

Let us stress that a specific dependence of probabilities on
the transversal coordinates $\xi~, \eta$ vanishes for a homogeneous
regular magnetic field which is a function of the longitudinal coordinate
$r = t$ only. Therefore for scenarios with regular fields (for
instance, in \cite{Akhmedov}) the spectrum Eq.  (\ref{spectrum})
takes the usual form without dependence on transversal coordinates.

If we assume uniform partial neutrino fluxes,
the  integration over the transversal coordinates $\xi,~\eta$ in the
integrand of the event number Eq.  (\ref{spectrum}) leads to the
averaging of  probabilities, see   Eq. (\ref{factor}) and
subsection III.D.

The energy resolution in the SK experiment is given by $\Delta (T) =
\Delta_{10}\sqrt{T/10~MeV}$ and  for the Kamiokande
experiment is estimated as $\Delta T/T\sim 0.2$ at the energy $T = 10~MeV$,
or $\Delta_{10} = 2~MeV$. This irreducible systematical error becomes
even worse near the threshold ($T_{th} = 6.5~MeV$ at the present time
and one plans to reach $T_{th}= 5~MeV$ in 1998).

In BOREXINO (starts in 1999), where one has a liquid scintillator, it
is expected to observe approximately 300 photoelectrons ({\it phe})
per $MeV$ of deposited recoil electron energy. This gives an estimate
of the energy resolution of the order\cite{BOREXINO}
\begin{equation}
\frac{\Delta T}{T}(BOREXINO) \simeq \frac{1}{\sqrt{N_{phe}}}\simeq
\frac{0.058}{\sqrt{T/MeV}}~,
\label{borex}
\end{equation}
corresponding to 12 \% for the threshold ($T_{th}\simeq 0.25~MeV$)
and 7 \% for the maximum recoil energy for Be neutrinos,
$T_{max}\simeq
0.663~MeV$.

In HELLAZ the Multi-Wire-Chamber (MWC) counts the secondary
electrons produced by the initial one in helium. It is expected to
count 2500 electrons at the threshold energy $T_{th}\simeq 0.1~MeV$,
so
in that case\cite{HELLAZ}
\begin{equation}
\frac{\Delta T}{T}(HELLAZ) \simeq \frac{1}{\sqrt{N_e}}\simeq
\frac{0.02}{\sqrt{T/T_{th}}}~,
\label{hellaz}
\end{equation}
or the energy resolution of the order 1.5 \% for the maximum for $pp$
neutrinos $T_{max}\simeq 0.26~MeV$.

Other experimental uncertainties must be incorporated into the
differential spectra Eq. (\ref{spectrum}). We have neglected an
unknown
statistical error, which we expect to be less than the systematical
one after enough exposition time, as it will decrease as $\mid \pm \Delta
N_{\nu}/N_{\nu}\mid \sim t^{-1/2}$. We have also neglected inner and
external background contributions to the systematical error considering
them as the specifical ones for each experiment.

\subsection{Radiochemical experiments}
The number of neutrino events in GALLEX (SAGE) and Homestake experiments
measured in SNU (1 SNU = $10^{-36}$ captures per atom/per sec) has
the form

\begin{equation}
\label{radio}
10^{-36} N^{\nu}_{Ga,~Cl} =
\sum_i\Phi^{(0)}_i\int_{E_{th}(Ga,Cl)}^{E_{max}(i)}
\lambda_i(E)\sigma_{Ga,~Cl}(E) <P_{ee}(E)>\, dE~,
\end{equation}
\noindent
where the thresholds $E_{th}(Ga,Cl)$ for GALLEX (SAGE) and Homestake are
$0.233~MeV$ and $0.8~MeV$ correspondingly and the capture cross sections
$\sigma_{Ga,~Cl}(E)$ for gallium and clorine detectors are tabulated in
\cite{Bahcall}, see Table 8.4.

For $P_{ee}=1$ the SSM predictions for radiochemical experiments are
listed in the Table 8.2.  \cite{Bahcall}.
For updated theoretical fluxes in BP95 model\cite{BP}
we find the mean SSM predictions
137 SNU for GALLEX (SAGE) and 9,3 SNU for the
Homestake experiment. The experimental data\cite{radiochem} pronounce the
electron neutrino deficit, $69.7 \pm 6.7^{+3.9}_{-4.5~}$SNU for the GALLEX,
$69 \pm 10^{+5}_{-7~}$SNU for the SAGE and $2.55 \pm 0.17\pm 0.18~$SNU for the
Homestake experiments.

Notice that the ratio of the experimental data to SSM predictions above
does not depend on theoretical uncertainties of the integral neutrino fluxes,
$\Phi^{(0)}_i$, in Eq. (\ref{radio}). Neutrino deficit expressed through
these ratios\cite{Berezinsky} is $0.509\pm 0.059$, $0.504 \pm 0.085$ and
$0.274\pm 0.027$ for GALLEX, SAGE and Homestake correspondingly where the
experimental 1$\sigma$-errors come from
$\sqrt{\sigma^2_{stat} + \sigma^2_{sys}}$.
\section{Master equation and its solution}
\subsection{Master equation}

We consider conversions $\nu_{eL,R}\to \nu_{aL,R}$, $a = \mu$ or
$\tau$, for two neutrino flavors obeying the master evolution
equation

\widetext
\begin{equation}
i\left(
\begin{array}{l}
\dot{\nu}_{eL}\\
\dot{\tilde{\nu}}_{eR}\\
\dot{\nu}_{\mu L}\\
\dot{\tilde{\nu}}_{\mu R}
\end{array}
\right) =
\left(
\begin{array}{cccc}
V_e -c_2\delta & 0 & s_2\delta & \mu H_+(t) \\
0 & - V_e - c_2\delta & - \mu H_-(t)& s_2\delta \\
s_2\delta & - \mu H_+(t)&V_{\mu} + c_2\delta &0 \\
\mu H_-(t)&s_2\delta &0& - V_{\mu} + c_2\delta
\end{array}
\right)
\left(
\begin{array}{c}
\nu_{eL}\\
\tilde{\nu}_{eR}\\
\nu_{\mu L} \\
\tilde{\nu}_{\mu R}
\end{array}
\right)~,
\label{master}
\end{equation}
\narrowtext
\noindent
where $c_2 = \cos 2\theta$, $s_2 = \sin 2\theta$, $\delta = \Delta
m^2/4E$ are the neutrino mixing parameters; $\mu = \mu_{12}$ is the
neutrino active-active transition magnetic moment;
$H_{\pm}(t) = B_{0\pm}(t) + b_{\pm}(t)$, $H_{\pm} = H_x \pm
iH_y$, are the magnetic field components consisting of regular
($B_{0\pm}$) and random ($b_{\pm}$) parts which are
perpendicular to the neutrino trajectory in the Sun; $V_e(t) =
G_F\sqrt{2}(\rho (t)/m_p)(Y_e - Y_n/2)$ and $V_{\mu}(t) =
G_F\sqrt{2}(\rho (t)/m_p)(- Y_n/2)$ are the neutrino vector
potentials
for $\nu_{eL}$ and $\nu_{\mu L}$ in the Sun given by the abundances
of
the electron ($Y_e = m_pN_e(t)/\rho (t)$) and neutron ($Y_n =
m_pN_n(t)/\rho (t)$) components and by the SSM density profile $\rho
(t) = 250~g/cm^{3~}\exp ( - 10.54t)$ \cite{Bahcall}.

Notice that we did not assume here a twist field model
with analogous constructions $B_{\pm} = B_x \pm iB_y$ in
off-diagonal entries of the Hamiltonian in Eq. (\ref{master})
These expressions are derived from the initial Majorana equations
in the mass-eigenstate representation where transversal part of the
spin-flip term $\mu \vec{\sigma}\cdot \vec{B}$
leads automatically to the "twist-form" in the Schr\"{o}dinger
equation above written in flavor representation (see \cite{Grimus}).
We have not performed the phase transform of the Hamiltonian in
Eq. (\ref{master}) since, in contrast to
\cite{APS}, the phase $\Phi (t)$ in
$H_{\pm} = H_{\perp}(t)e^{\pm i\Phi (t)}$ is the random one that
gives uncertainty  in the additional term $\dot{\Phi}$ for the
resonance condition \cite{APS}. Instead of that we have solved
Eq. (\ref{master}) directly using computer simulation (see below).
\subsection{Solar magnetic fields}
\subsubsection{Random magnetic fields}

The r.m.s. random component {\bf b} is assumed to be stronger then
the regular one, {\bf B}, and maybe even much stronger than
$B_0$, $\langle b^2\rangle \gg B_0^2$, provided the large magnetic
Reynolds number $R_m$ leads to the effective dynamo enhancement of
small-scale (random) magnetic fields.

Let us give simple estimates of the magnetic Reynolds number $R_m =
lv/\nu_m$
in the convective zone for fully ionized hydrogen plasma ($T\gg
I_H\sim 13.6~eV\sim 10^5~K$). Here $l\sim 10^8~cm$ is the size of
eddy (of the order of magnitude of a granule size) with the turbulent
velocity inside of it $v\sim 10^5~cm/s$. Provided an equipartition
between the turbulent kinetic energy and the {\it mean} magnetic
field is suggested, we obtain $v \approx v_A = B/\sqrt{4\pi \rho}$,
where $v_A$ is the Alfven velocity for MHD plasma, $B$ is the mean
(large-scale) magnetic field in convective zone and $\rho$ is the
matter density The magnetic diffusion coefficient, or magnetic
viscosity, $\nu_m = c^2/4\pi \sigma_{cond}$ entering the
diffusion term  of the Faradey induction equation,

\begin{equation}
\frac{\partial {\bf \vec{H}}}{\partial t} =
rot [{\bf \vec{v}}\times {\bf \vec{H}}]
+ \nu_m \Delta {\bf \vec{H}}~,
\label{Faradey}
\end{equation}
where ${\bf \vec{H}} = {\bf \vec{B}}+{\bf \vec{b}}$ is the total
magnetic field (mean field plus fluctuations), is given by the
conductivity of the hydrogen plasma
$\sigma_{cond} = \omega_{pl}^2/4\pi \nu_{ep}$. Here $c$ is the
light velocity; $\omega_{pl} = \sqrt{4\pi e^2n_e/m_e} = 5.65\times
10^4\sqrt{n_e}~s^ {-1}$ is the plasma, or Lengmuir frequency; $\nu_{ep}
= 50n_e/T^{3/2}~s^{-1}$ is the electron-proton collision frequency; and
the electron density $n_e = n_p$ and the temperature $T$ are measured in
$cm^{-3}$ and  Kelvin degrees respectively.

It follows that  the magnetic diffusion coefficient
$\nu_m\simeq 10^{13}(T/1~K)^{-3/2}~cm^2s^{-1}$ does not depend
on the charge density $n_e$ and it is very small
for hot plasma $T\geq 10^5~K$. From comparison of the first and
second terms in the r.h.s. of the Faradey equation Eq. (\ref{Faradey})
we find that $v/l\gg \nu_m/l^2$, or $\nu_m\ll
vl\sim 10^{13}~cm^2s^{-1}$ since $T/1~K\gg 1$.
This means that magnetic field in the Sun is {\it frozen-in}.

As the result we obtain $R_m =
lv\omega_{pl}^2/(c^2\nu_{ep})\simeq
(10^{-13}lv/ cm^2 s^{-1})(T/1~K)^{3/2}$.
A standard estimation for $R_m$ in the solar convective zone is
$R_m = 10^8$ (see \cite{Zeldovich}).

The estimation of $b/B$ for the solar convective zone (and other
cosmic dynamos) is the matter of current scientific discussions.
The most concervative estimate, simply based on equipartition concept,
is $b = const \, B$. According to  direct observations of galactic
magnetic field presumably driven by a dynamo, $b \approx 1.8 B$
(\cite{Ruzmaikin}). A more developed theory of equipartition
gives, say, $b = const B \ln R_m$ (see \cite{Zeldovich}). For
our consideration taking this constant as $4 \pi$
would be more than sufficient.

Notice that this estimate is considered now as
very concervative. Basing on more detailed theories of MHD turbulence
estimates like $b \sim \sqrt{R_m} B$
are discussed \cite{Cataneo}, \cite{Diamond}. Such estimates give a free
room for a very large values of $b$. Let us stress, however, that the
larger estimate of $b$ is accepted, the more difficult is to refer for a
dynamo as an origin for $B$, so the estimates of Vainstein and Cataneo,
\cite{Cataneo} and \cite{Diamond} are hardly comparable with the
dynamo nature of large-scale solar magnetic field.

Being interested in neutrino conversions, we need only a partial
information concerning the small-scale magnetic fields. In particular,
because of the very rapid propagation of neutrino in comparison with
MHD timescales, we are interesting only in a $b$ distribution at a given
instant in a given direction.
\subsubsection{The model of the magnetic field}

The  assumed profile of a random magnetic field is presented in the
Fig.\ \ref{fig02}.

We also show there the profile of the regular magnetic field
used in calculations.
We find that RSFP results are not sensitive to the profile of a regular
field.  However, switching on random fields,
we find out an essential change of the probabilities $P_{aa}(t)$ for
the same mixing parameters $\sin^22\theta$, $\Delta m^2/2E$. This
difference of the magnetic field model influence on the solution of the
master equation Eq. (\ref{master}) becomes the more significant the
larger the r.m.s. value $b$ is assumed.

The numerical implementation for the random magnetic field has been
chosen as follows. We choose the correlation
length for the random field component as $L_0=10^4~km$ that is close to
the mesogranule size.

Then we suppose that
all the volume of the convective zone is covered by a net
of rectangular domains where the random magnetic field
strength vector is constant. The magnetic field strength changes
smoothly at the boundaries between the neighbour domains obeying the
Maxwell equations. Since one can not expect the strong influence of
small details in the random magnetic field within and near thin boundary
layers between the domains this oversimplified model looks applicable.

In agreement with the SSM \cite{Bahcall} we
suppose that the neutrino source inside the Sun is located inside the
core with the radius of the order of $R_{\nu} = 0.1R_{\odot}= 7\times
10^4~km$ and neglect for the sake of simplicity the spatial
distribution of neutrino emissivity from a unit of a solid angle
of the  core image.
However, different parallel trajectories directed to the
Earth cross different magnetic domains because the domain size $L_0$
(in the plane which is perpendicular to neutrino trajectories) is
much
less than the transversal size $=R_{\nu}$ of the full set of parallel
rays, $L_0\ll R_{\nu}$.
The whole number of  trajectories (rays) with statistically
independent magnetic fields is about $R_{\nu}^2/L_0^2\sim 50$.

Thus, the perpendicular components $b_{x,y}(r)$ along one
representive ray are the random functions with the correlation length
$L_0$ (see Fig.\ \ref{fig02}).

At the stage of the numerical simulation of the random magnetic field
we generate the set of the random numbers with the given r.m.s.
$\sqrt{\langle b^2\rangle}$ and suppose that the strength of the
random
field is constant inside the rectangular volume with the radial size
$L_0$ and with the same diameter $L_0$ in the $\xi~,\eta$-plane. We
assume that
$\sqrt{\langle b_x^2\rangle} = \sqrt{\langle b_y^2\rangle}$.

We generate
random values of $\bf b$ in different cells both along
the neutrino trajectory and in the transversal plane $\xi~,\eta$,
solve
the Cauchy problem for 50 rays and calculate the mean
 probabilities Eq. (\ref{averprob}).

Notice that we substitute
into the integrand of the neutrino event number $dN_{\nu}/dT$ the
probabilities $P_{aa}(r = R_{E\odot}, \xi, \eta)$ at the Earth taking
into account also vacuum neutrino oscillation probabilities (for zero
magnetic fields and zero density in the solar wind) and neglecting the
Earth effects for relatively small $\Delta m^2\sim 10^{-8}~eV^2$
($l_v =2.48\times 10^{-3}E(MeV)/\Delta m^2(eV^2)~km\gg l_0
= 3.28\times 10^4~km/\rho (g/cm^3)$\\
for $\rho\sim 4.5--11~g/cm^3$ in the Earth). As for MSW-region
$\Delta m^2\sim 10^{-5}~eV^2$ we do not consider here the possible
Earth influence which could result in day-night variations.

\subsection{Asymptotic solution of the master equation}

Let us consider how random magnetic field influences the small-mixing
MSW solution to SNP with the help of some simplified analytical solutions
of  the  master  equation  Eq. (\ref{master}).
Indeed, it turns out that for the SSM exponential density  profile,
typical borone neutrino energies
$E \sim 7\div14 MeV$,
and
$\triangle m^2\sim10^{-5}~eV^2$,
the MSW resonance
(i.e. the point where $V_e-V_{\mu}=2c_2\delta)$
occurs well below the bottom of the convective zone.
Thus  we  can  divide  the  neutrino
propagation problem and  consider  two  successive  stages.
First, after generation in the middle of the Sun,
neutrinos propagate in the absence of any magnetic  fields,
undergo  the non-adiabatic (non-complete) MSW  conversion
$\nu_{eL} \to \nu_{\mu L}$
and  acquire  certain  nonzero   values
$P_{ee}$  end  $P_{\mu\mu}$,
which  can  be  treated  as   initial conditions
at the bottom of the convective zone.
For small  neutrino mixing
$s_2\to 0$  the  $(4\times4)$-master  equation  Eq.(8) then splits
into  two  pairs  of  independent   equations
describing correspondingly  the   spin-flavor    dynamics
$\nu_{el} \to \tilde\nu_{\mu R}$
and
$\nu_{\mu L} \to  \tilde\nu_{eR}$
in  noisy magnetic fields.
In addition, once the MSW resonance point is far away
from the convective zone,
one can also omit $V_e$ and $V_{\mu}$ in comparison with $c_2\delta$.
For $\nu_{eL}\to \tilde\nu_{\mu R}$ conversion
this results into a  two-component  Schr$\rm \ddot o$dinger equation
\begin{equation}
  i\left(\begin{array}{c}
    \dot\nu_{eL}\\
    \dot{\tilde\nu}_{\mu R}
         \end{array}\right)
  \simeq \left(\begin{array}{cc}
   -\delta & \mu b_{+}(t)\\
    \mu b_{-}(t) & \delta
               \end{array}\right)
  \left(\begin{array}{c}
    \nu_{eL}\\
    \tilde\nu_{\mu R}
        \end{array}\right)
\label{(01)}
\end{equation}
with initial conditions
$
  |\nu_{eL}|^2(0)=P_{ee}(0),
  \quad
  |\tilde\nu_{\mu R}|^2 = P_{\tilde\mu\tilde\mu}(0) =0.
$
As normalized probabilities
$P_{ee}(t)$
and
$P_{\tilde\mu\tilde\mu}(t)$
(satisfying the conservation law
$P_{ee}(t) + P_{\tilde\mu \tilde\mu}(t) =  P_{ee}(0)$)
are the only observables,
it is convenient to recast the Eq.(\ref{(01)})
into an equivalent integral form
\begin{eqnarray}
  S(t) &=& S(0) - 4\mu^2
    \int \limits_0^t dt_1
    \int \limits_0^{t_{1}} dt_2
    S(t_2) \nonumber \\
& &\{[b_x(t_1)b_x(t_2) + b_y(t_1)b_y(t_2)] \cos2\delta(t_1 - t_2)
  \nonumber    \\
& &-[b_x(t_1)b_y(t_2) - b_x(t_2)b_y(t_1)] \sin2\delta(t_1 - t_2) \},
\label{(03)}
\end{eqnarray}
\noindent
where
$S(t)=2P_{ee}(t)-P_{ee}(0) \equiv P_{ee}(t)-P_{\tilde\mu\tilde\mu}(t)$
is the third component of the polarization vector $\vec S$.
\subsubsection{$\delta$-correlations}
If $L_0$ is the minimal physical scale in the  problem,
we can consider random magnetic fields as $\delta$-correlated:
\begin{eqnarray}
  <b_x(t_1) b_x(t_2) +b_y(t_1)b_y(t_2)> &=&
  \frac23 <{\vec b}^2> \cdot L_0\delta(t_1-t_2),
\nonumber\\
<b_xb_y> &=& 0.
\nonumber
\end{eqnarray}
In this case it turns out
that averaging of the Eq.(\ref{(03)}) over random magnetic fields
is exactly equivalent to averaging of its solution (cf. \cite{Enqvist}).
The result reads:
\begin{equation}
  <S(t)> = S(0) e^{-\Gamma t},
  \label{(04)}
\end{equation}
where $<S(t)>$ is the mean value, and decay factor
\begin{equation}
  \Gamma = \frac43 \mu^2 <\vec b^2>L_0.
  \label{(05)}
\end{equation}
If time is measured in units of $L_0$
(recall that  $c=1$),
supposing that neutrino traversed $N$ correlation cells,
i.e. $t=L_0 \times N$, we have
\begin{equation}
  <S(N)>  =  S(0)
  \exp \left\{ -\frac43\mu^2 <\vec b^2> L_0^2 \times N \right\},
  \label{(06)}
\end{equation}
so that N plays a role of an extensive variable
like the volume  $V$  in statistical  mechanics.
\subsubsection{Correlations of finite radius}
Let us consider Eq.(\ref{(03)}) in a spirit of our numerical method.
Dividing the interval of integration
into a set of  equal intervals
of correlation length $L_0,$ for a current cell $n$ we have
\begin{eqnarray}
& & S_{n} \equiv S(n \cdot L_0)= \nonumber  \\
& &  = S(0)-4\mu^2
  \sum\limits^n_{k=1} \sum\limits^k_{l=1}
  \int\limits^{kL_0}_{(k-1)L_0} dt_1
  \int\limits^{\min \{t_1,lL_0 \} }_{(l-1)L_0} dt_2
  \; \{ \mbox {the integrand  of Eq.(\ref{(03)})} \}.
  \label{(07)}
\end{eqnarray}
Assuming    now    that    possible     correlations        between
$S(t_2)$ and $b_i , b_j$ under the integral are  small,
$S(t)$
itself varies very slowly within one correlation cell
(see,  however,  below),
and  making  use  of    statistical properties
of random fields,
($i$) different  transversal  components
within one  cell  are  independent  random  variables,  and
($ii$)
magnetic fields in different cells do not correlate,
we can average Eq.(\ref{(07)})
thus obtaining a finite difference analogue:
\begin{eqnarray}
& & S_{n} = S(0)
  -4\mu^2\sum\limits^n_{k=1}
  <\vec b_{\perp k}^2>
  S_{k}
  \int\limits^{kL_0}_{(k-1)L_0} dt_1
  \int\limits^{t_1}_{(k-1)L_0}  dt_2
  \cos 2\delta(t_1 - t_2)= \nonumber \\
& & = S(0) - 2\mu^2 \frac{\sin^2\delta L_0}{\delta^2}
  \sum\limits^n_{k=1} <\vec b^2_{\perp k}> S_{k},
  \label{(08)}
\end{eqnarray}
where we retained possible slow space  dependence
of the  r.m.s. magnetic field value
$\displaystyle<\vec  b_{\perp  k}^2> = \frac23<\vec  b_k^2>$.
Returning now to continuous version of Eq.(\ref{(08)}) we get
\begin{equation}
  S(t) = -\frac43\mu^2L_0
    \frac{\sin^2\delta L_0}{(\delta L_0)^2}
    \int\limits^t_0dt' <\vec b^2(t')> S(t'),
\end{equation}
the solution of which has a form
\begin{equation}
  S(t) = S(0) \exp \left\{ -\frac43 \mu^2L_0
    \frac{\sin^2\delta  L_0}{(\delta  L_0)^2}
    \int\limits^t_0 <\vec b^2(t')> dt'\right\}.
\label{finite}
\end{equation}
For $\delta \ll L_0^{-1}$
and  constant  r.m.s.  $<\vec b^2>$
we  obtain  the  simple  $\delta$-correlation  result
Eq.(\ref{(04)}).
The last form, however, allows to evaluate at what part of
the convective zone the ASFC really takes place for a given profile
of the r.m.s. $<\vec b^2>$.
\subsubsection{Linked cluster expansion and dispersion}
Another important issue
is the problem of temporal dependence of higher statistical moments
of $P_{aa}.$
As $P_{aa}$ enter the Eq.(\ref{integral}) for the number of events
one should be certain that the averaging procedure
does not input large statistical errors,
otherwise there will be no room
for the solar neutrino puzzle itself.
In order to apparently investigate
generic statistical properties of the model
we make an additional assumption
and put in Eqs.(\ref{(01)}),(\ref{(03)})
$\delta$ equal to zero.
This is not critical at least for strong random magnetic fields
as it follows that for small $\delta$
instantaneous eigenvalues of Eq.(\ref{(01)})
\begin{equation}
  E_{\pm}(t) =
  \pm \sqrt{\delta^2 + \mu^2 {\vec b_\perp}^2(t)}
  \approx \pm \mu |{\vec b_\perp}(t)| ,
\end{equation}
i.e. for typical realization the vacuum oscillation parameter
should not substantially influence the solution provided also
that $\delta \ll L_0^{-1}$ (see Eq. ( \ref{finite}) ).
The integral equation (\ref{(03)}) then takes the form
\begin{equation}
  S(t) = S(0) - 4\mu^2
    \int \limits_0^t dt_1
    \int \limits_0^{t_{1}} dt_2
    {\vec b}_{\perp}(t_1){\vec b}_{\perp}(t_2) S(t_2),
\label{(C1)}
\end{equation}
especially convenient for perturbative solution
in powers of magnetic field or, equally,
in powers of coupling constant $4\mu^2.$

The $n$-th order is proportional to the product of $2n$
transversal components ${\vec b}_{\perp}(t_i)$
integrated over time with descending upper limits of integration
$t>t_1>t_2>\dots>t_{2n-1}.$
After averaging and some standard combinatorics
(cf. \cite{Abrikosov})
the perturbation series exponentiates
providing the linked cluster  expansion
\begin{equation}
  <S(t)> = S(0) \exp
  \left\{
    \sum\limits_{m=1}^{\infty} \frac{(-4\mu^2)^m}{(2m)!} M_{2m}(t)
  \right\},
  \label{(C2)}
\end{equation}
where because of isotropy we retained
linked higher moments of even order only,
which can be evaluated within our spacially piece-constant model
of random magnetic fields as
\begin{equation}
  M_{2m}(t) \simeq \frac{t}{L_0}
  \left( <{\vec b}_{\perp}^{2}> L_0^2 \right)^m \lambda_{2m},
  \label{(C3)}
\end{equation}
where $\lambda_{2m}$ -- some coefficients,
$0 \leq \lambda_{2m} \leq 1, \; \lambda_2=1,$
defined by magnetic field statistics within one cell.
For Gaussian distribution all $\lambda_{2m}$ but $\lambda_2$
disappear and Eq.(\ref{(C2)}) coincides with the $\delta$-correlator
result Eq.(\ref{(04)}).
Approximately this is also true when
$\Omega \equiv \mu L_0 \cdot \sqrt{ <{\vec b}_{\perp}^{2}> } \ll 1$
due to fast convergence of the sum in Eq.(\ref{(C2)}).
For chosen values of magnetic fields and correlation scale
this parameter does not exceed $\sim 0.2,$
thus justifying the $\delta$ - correlation estimates.
It is also plausible (though we have not proved that correctly)
that once we adopted the cell model
with physically constrained from above random magnetic fields,
the resulting approach of $<S(t)>$ to its asymptotic value
should be always exponential
leaving no room to such effects as intermittency.
That is $<S(t)>$ is simply decaying with time tending to zero.
Here we illustrate such a behavior
for a type of statistics  differing from Gaussian.

{\it Stochastic twist}.
Let transversal random magnetic fields
in every correlation cell are constant {\it by modulo},
$<{\vec b}_{\perp k}^{2}> = <{\vec b}_{\perp}^{2}> = {\rm const}$ ,
differing from each other only by random direction of vector
${\vec b}_{\perp k}.$
Then all even orders of the field entering
Eqs.(\ref{(C2)}), (\ref{(C3)})
are constant too and $\lambda_{2m}=1.$
The sum in Eq.(\ref{(C2)}) is easily performed and we obtain
\begin{equation}
  <S(t)> = S(0) \exp
  \left\{
    - \frac{2t}{L_0}\sin^2 \Omega
  \right\},
  \label{(C4)}
\end{equation}
where $\Omega = \mu |{\vec b}_\perp| L_0.$
We note by passing that for large magnetic fields satisfying
rather artificial conditions
$\Omega=\pi n, \; n=1,2,\dots,$
there should exist an effect of resonance transparency,
when the polarization vector $\vec S$ performs
an integer or half-integer number of turnovers within one cell.
Otherwise the behaviour is exponential again.

\vskip 0.5cm

To estimate the $<S^2(t)>$ behaviour we use the same procedure,
write down the iterative solution for $S(t),$
square it and then average.
The final result again has an exponential form like ( \ref{(C2)}).
As it turns out that $\Omega \ll 1$
we confine here only with the Gaussian statistics
(valid in this case).
Then,
\begin{equation}
  <S(t)> = S(0) e^{-\Gamma t},
  \quad
  <S^2(t)> = \frac{S^2(0)}{2} \left(1+ e^{-4\Gamma t}\right),
  \label{(C5)}
\end{equation}
where $\Gamma$ is defined in Eq.(\ref{(05)})
and we also rewrite the expression (\ref{(04)}) for ${<S(t)>.}$
It follows from Eq.(\ref{(C5)}) that
$<S^2(0)>=S^2(0),$ and for $t\to\infty$ the exponent dies out
and $<S^2(t)>$ tends to its asymptotic value
$<S^2(\infty)>=\frac12S^2(0).$
For dispersion $\sigma^2_S$ we then obtain
\begin{equation}
  \sigma_S = \sqrt{<S^2(t)> - <S(t)>^2}
  = \frac{S(0)}{\sqrt2} \left( 1 -  e^{- 2 \Gamma t} \right),
  \label{(C7)}
\end{equation}
and correspondingly for $\sigma_P$
\begin{equation}
  \sigma_P = \sqrt{<P_{ee}^2(t)> - <P_{ee}(t)>^2}
  = \frac{P_{ee}(0)}{2\sqrt2} \left( 1 -  e^{- 2 \Gamma t} \right).
  \label{(C8)}
\end{equation}
Taking into  account that
\begin{equation}
  <P_{ee}(t)> = \frac{<P_{ee}(0)>}{2} \left( 1 + e^{- \Gamma t} \right),
  \label{(C9)}
\end{equation}
we have that relative mean square deviation of $P_{ee}$
from its mean value tends with $t\to\infty$
to its maximum asymptotic value
\begin{equation}
  \frac{\sigma_P(t)}{<P_{ee}(t)>}
  \to \frac{1}{\sqrt2} \simeq 0.707 , \qquad t \to \infty,
  \label{(C10)}
\end{equation}
irrespectively of the initial value $P_{ee}(0).$
It is interesting that this asymptotic value is in a qualitative
agreement with the result \cite{Balantekin},
\begin{equation}
   \frac{\sigma_P(t)}{<P_{ee}(t)>}
   \simeq 0.58 , \qquad t \to \infty,
   \label{(C11)}
\end{equation}
that was obtained by numerical simulation of the MSW-effect
in a fluctuating matter density.

\vskip 0.5cm

We can now estimate the influence of the random magnetic field
on the process of the neutrino conversion.
Let suppose that the effective thickness
of the part of the convective zone of the Sun
carrying the large random magnetic field is about
$\Delta t \approx 0.15R_{\odot}$
and  $\mu = 10^{-11}\mu_B.$
Then $\Gamma \approx 1.6\times{b}/{100~kG}, \quad s^{-1}.$
Hence, random magnetic field with the strength about $100~kG$
is strong, mixing parameter
$K=1- e^{-\Gamma \Delta t} \approx 0.8.$
Random magnetic field $b\sim 50~kG$ is medium,
$K \approx 0.33$
and $b\sim 20~kG$ is weak,
$K \approx 0.06.$
These estimations are the reasons
to adopt the strength of the random magnetic field
for the computer simulation, $b=50~kG$ and $b=100~kG.$

The estimation (\ref{(C10)}) of the r.m.s. deviation of $P_{ee}$
and other $P_{aa}$ is true, evidently, only for one neutrino ray.
Averaging over $N$ independent rays lowers the value (\ref{(C10)})
in $\sqrt N$ times.
That is for our case of $N \approx 50$ rays we get that
maximum relative error should not exceed approximately $10\%,$
thus justifying the validity of our approach.
For smaller magnetic fields the situation is always better.

To conclude this section, it is neccesary to repeat that the above
estimate Eq. (\ref{(C10)}) indicates possible danger when treating
numerically the neutrino propagation in noisy media. Indeed,
usually adopted one-dimensional (i.e. along one ray only)
approximation for the $(4 \times 4)$ master equation Eq. (\ref{master})
or $(2 \times 2)$ Eq. (\ref{(01)}) can suffer from large dispersion
errors and one should make certain precautions when averaging these
equations over the random noise {\it before} numerical simulations.
Otherwise, the resulting error might be even unpredictable.

\subsubsection{Interpretation of the influence
of random magnetic field as a random walk over a circle}

Here we show that
there exists a simple way to explain the behaviour
of the mean value and dispersion of $P_{ee}(t).$
To apparently account the normalization
$P_{ee}(t) + P_{\tilde\mu \tilde\mu}(t) =  P_{ee}(0),$
we can introduce a unit circle where a representative point
parametrized by the angular variable $\phi(t)$ performs a motion,
while $P_{ee}(t)=P_{ee}(0)\cos^2 \phi(t),$
$P_{\tilde\mu \tilde\mu}(t)=P_{ee}(0)\sin^2 \phi(t).$
Then  expressions for $S$ and $S^2$ take form
$S(t) \equiv P_{ee}(t)-P_{\tilde\mu \tilde\mu}(t)=S(0)\cos 2 \phi,$
$\displaystyle S^2(t)=\frac{S^2(0)}{2} \left\{ 1+\cos 4 \phi \right\}.$
Suppose now that $\phi(t)$ is a realization of a Gaussian random process
with the dispersion $D_{\phi}(t)=\beta t$ and probability density
$$
  P(\phi) =\frac{1}{\sqrt{2\pi\beta t}}
  \exp\left\{-\frac{\phi^2}{2\beta t} \right\}.
$$
Simple calculation shows that
\begin{equation}
  <\cos \alpha \phi>=\exp\left\{- \frac{\alpha^2\beta t}{2} \right\}.
  \label{avecos}
\end{equation}
From Eq.(\ref{avecos}) it follows that
\begin{eqnarray}
  <S(t)> &=& S(0) e^{- 2\beta t },
  \nonumber\\
  <S^2(t)> &=& \frac{S^2(0)}{2} \left( 1 + e^{-8\beta t} \right),
  \nonumber\\
  \sigma_S &=& \frac{S(0)}{\sqrt2} \left( 1 - e^{-4\beta t} \right).
  \label{aves}
\end{eqnarray}
Comparison of  Eqs.(\ref{(C5)}), (\ref{(C7)})
with (\ref{aves}) shows that if we adopt
$2 \beta = \Gamma,$
these expressions become identical.
Hence, one can treat the master equation (\ref{(01)})
with the random magnetic field as a random walk {\it over a circle}
with the dispersion of the angular variable
proportional to the product of the path along the convective zone
$\delta t,$
mean squared magnetic field
$<{\vec b}_\perp^2>$
and correlation length $L_0,$
coefficient of proportionality
depending on the neutrino magnetic moment squared, see Eq.(\ref{(05)}).
This provides a simple way to express the meaning of the Eq.(\ref{(C9)}).
At the starting moment $\phi(0)$
possess probability density concentrated at point $\phi=0,$
and $P_{ee}(0)$ has the definite value governed by the initial conditions.
If $t \to \infty,$ the random walk over a circle
provides an asymptotically uniform probability density,
$P_{\phi}(t=+\infty)={1}/{2\pi}$
and
$<P_{ee}(t=+\infty)>=P_{ee}(0)/2$
because the average value of $\cos^2\phi$ is equal to $1/2.$
A more detailed study of possible  interrelation
between  neutrino conversions and the random walk will be reported
elsewhere, here we only add that an explicit computation of the
third and fourth moments of $S$ confirm the above identification.
Our results also confirm the conjecture \cite{Nicolaidis} where with
account of only one component of the transversal magnetic field
(scalar case) it was shown that the resulting behaviour of $<P_{ee}>$
can be interpreted as a brownian motion of an auxiliary angular
variable over a circle.

\subsection{Computer simulation  of the master equation}

Substituting a random realization of magnetic fields along one ray
(see curve 3 in Fig.\ \ref{fig02} for $b_x(r)$~\footnote{In general, for
any random field realization the component $b_y(r)$ has a different
profile (along r) shown in Fig.\ \ref{fig02} for $b_x(r)$ only. This difference
in profiles was taken into account while equal amplitudes,
$\sqrt{\langle b_x^2\rangle} = \sqrt{\langle
b_y^2\rangle}=\sqrt{\langle b^2\rangle}/3 $ , were assumed too.}) into
the master equation Eq.  (\ref{master}) we find the solution for four
wave functions, $\nu_a(t) = \mid \nu_a(t)\mid \exp (i\alpha_a(t))$,
from which all dynamical probabibilities $P_{aa}$ obeying the condition
Eq.  (\ref{unitarity}) are derived. Then we have repeated such
procedure with the solution of the Cauchy problem for other
configurations of random magnetic fields (along other neutrino
trajectories).

After that, supposing that due to homogeneity the
intensities of partial neutrino fluxes are equal to each other, i.e.
$\Phi_i^{(0)}(\xi , \eta) = \Phi^{(0)}_i=const$~, we obtain the number
event spectrum Eq.  (\ref{spectrum}) that depends on the product\\
$[\int d\xi d\eta]^{-1} \int d\xi d\eta \Phi_i^{(0)}(\xi ,
\eta)P_{aa}(\xi~, \eta) = \Phi^{(0)}\times \langle P_{aa}\rangle$.

Here $\Phi^{(0)}_i$ is
the integral flux of neutrinos of the kind "i" \cite{Bahcall} and
$\langle P_{aa}\rangle$ are the
mean probabilities that are shown in Figures\ \ref{fig03}--\ref{fig06}
and in Figure\ \ref{fig07}
for particular cases of the maximum r.m.s. amplitudes $\sqrt{\langle b^2
\rangle} = 50~kG$ and $\sqrt{\langle b^2
\rangle} = 100~kG$ correspondingly.

For comparison of ASFC with the RSFP case we substituted regular
magnetic fields $B_0 =50~,100~kG$ (with profiles shown in the same
Fig.\ \ref{fig02}~) into Eq.  (\ref{master}) and calculated four probabilities
$P_{aa}$.  The transition probabilities for antineutrinos
$P_{\bar{e}\bar{e}}$ (appeared for large mixing angles in cascade
conversions through the steps: (a) $\nu_{eL}\to \bar{\nu}_{\mu R}$
within convective zone and (b) $\bar{\nu}_{\mu R}\to \bar{\nu}_{eR}$ in
solar wind via vacuum conversions) are shown for the cases
$B=50~kG$ and $B = 100~kG$ in the Figures\ \ref{fig08},\ref{fig09}.


In Figures\ \ref{fig03}--\ref{fig09} along y-axis we plotted the
dimensionless parameter
$\delta = 8.81\times 10^8\Delta m^2(eV^2)/E(MeV)$.
For instance, for boron
neutrino energy $E\sim 8.8~MeV$ and $\Delta m^2 = 10^{-8}~eV^2$ we
find $\delta\sim 1$.

All four probabilities in the cases Figs.\ \ref{fig03}-\ref{fig07}
obey the unitarity condition Eq.  (\ref{unitarity}) for
the same parameters $\sin^22\theta$, $\delta = \Delta m^2/4E$ and for
a given $b_{rms}$. This can be viewed as a check of our numerical
procedure because the unitarity condition was not apparently used in
simulation.

In Figs.\ \ref{fig10}, \ref{fig11}
we plot allowed parameter regions $\Delta m^2~,
\sin^22\theta$ found from reconcilement of all experimental data
\cite{SK}, \cite{radiochem} for the ratio $DATA/SSM~\pm
1\sigma$ in the case of regular fields $B_{\perp}=50,~100~kG$ and in
Figs.\ \ref{fig12}, \ref{fig13}
we present analogous results for random fields
$\sqrt{\langle b^2\rangle} = 50,~100~kG$. In order to find these regions
we have calculated number of events in Eq. (\ref{integral}) using
corresponding numerical solutions of the evolution equation Eq.
(\ref{master}) with varying $\Delta m^2$,~$\sin^22\theta$ devided
by the number of events derived in SSM for $P_{ee}(E) = 1$.


In Figs.\ \ref{fig10}-\ref{fig14} we also show  $\Delta m^2$,~$\sin^22\theta$- regions
excluded from non-observation of antineutrinos
$\bar{\nu}_{eR}$ in SK \cite{Fiorentini}.

\vskip 0.3cm
\section {Discussion of results}

From Figs.\ \ref{fig03}--\ref{fig13} it apparently follows that
there is a band in the parameter
region $\Delta m^2\sim 10^{-7}-10^{-6}~eV^2$ that separates MSW
and magnetic field scenarios. In Figs.\ \ref{fig12}--\ref{fig14}
we presented the allowed
parameter regions to reconcile four solar neutrino experiments along
with the SK bound on $\bar{\nu}_{eR}$. One can see that without the
latter bound there exist two commonly adopted parameter regions,
the small mixing and the large mixing ones, currently allowed
(Figs.\ \ref{fig12},\ref{fig13})
or excluded (Fig.\ \ref{fig14}) in dependence of the values of the
r.m.s. magnetic field parameters, see subsections IV.B and IV.C. But
first of all we would like to discuss a "strange" dependence of the
allowed from the SK experiment low mass difference regions
($\Delta m^2\lsim 10^{-8}~eV^2$) on the strength of the regular
magnetic field.

\subsection{"Paradox" of strong regular magnetic
fields}

We find a surprising from the first sight result
for the strongest field $B = 100~kG$:  (i) the probability
$P_{\bar{e}\bar{e}}$ decreases relative to the case $B= 50~kG$ (compare
Figs. 8,9);  moreover, (ii) the well-known parameter region $\Delta
m^2\sim 10^{-8}~eV^2$ allowed here from SK data in the case of regular
field $B = 50~kG$ (shown in Fig.\ \ref{fig10}) and recoinciled with other
experiments in \cite{Akhmedov1} vanishes in the case of more stronger
field $B = 100~kG$ (see Fig.\ \ref{fig11}).

One can easily explain such peculiarity considering the analytic form of
neutrino conversion probability in the convective zone
($\nu_{eL}\to \bar{\nu}_{\mu R}$),
\begin{equation}
\label{analytic}
 P_{e\bar{\mu}} = \frac{(2\mu B_{\perp})^2}{(V - \Delta
\cos 2\theta)^2 + (2\mu B_{\perp})^2}
 \sin^2
\left(\sqrt{(V - \Delta \cos
2\theta)^2 + (2\mu B_{\perp})^2}\frac{\Delta r}{2}\right)~.
\end{equation}
This expression is valid for constant profiles and can be used to
qualitatively estimate the effect treated here numerically for changing
profiles. Here $\Delta r = 0.15R_{\odot}$ is the effective half-width of
the convective zone where magnetic field is strong (see Fig.\ \ref{fig02});
$V = \sqrt{2}G_F(\rho (r)/m_p)(Y_e - Y_n) \leq 1.6\times 10^{-14}(Y_e -
Y_n)~eV$ is the neutrino vector potential ($V= V_e - V_{\bar{\mu}}$)
above the bottom of convective zone $r\geq 0.7R_{\odot}$ (see
Eq.  (\ref{master})); $2\mu B_{\perp}\simeq 5\times 10^{-15}B_{50}~eV$
is the magnetic field parameter for $\mu = 10^{-11}\mu_B$ and magnetic
field strength normalized on 50~kG.

The mass parameter $\Delta = 10^{-11}\Delta m^2_5/2(E/MeV)~eV$ is much
less than parameters above
for SK energies $E\geq 6.5~MeV$ in the mass parameter region $\Delta
m^2\leq 10^{-8}~eV^2$, or for $\Delta m^2_5\leq 10^{-3}$.

Since $\Delta m^2 (eV^2) = 4\times 10^{-9}-10^{-10}$ is negligible and
the resonance condition $V = \Delta \cos 2\theta$, i.e.,
\begin{equation}
\frac{\Delta m^2_5cos 2\theta}{2E/MeV} = 1.6(Y_e -
Y_n)\exp (-10.54r/R_{\odot})~,
\label{resonance}
\end{equation}
 is not fulfilled for corresponding low mass
region, we conclude and check
directly from the analytic formula Eq. (\ref{analytic}) that
in the case $B_{\perp} = 50~kG$ the
{\it non-resonant} large conversion probability $P_{e\bar{\mu}}$ reaches
 a maximum due to accident coincidence of the phase in propagating factor
with $\sqrt{V^2 + (2\mu B_{\perp})^2}\Delta r/2\sim \pi/2$.

This is in contrast to another allowed region $\Delta
m^2\sim 10^{-7}~eV^2$ shown in the same Fig.\ \ref{fig10} where really the
resonance Eq.  (\ref{resonance}) and RSFP take place.

Notice that the oscillation depth $(2\mu B_{\perp})^2/((V - \Delta
\cos 2\theta)^2 + (2\mu B_{\perp})^2)$  reaches
unity in Eq. (\ref{analytic}) for extreme fields
$B_{\perp}\to \infty$ and the resonance $V =
\Delta \cos 2\theta$ (\ref{resonance}) is almost irrelevant to the case
$B_{\perp} = 100~kG$. Similar behavior of RSFP was studied also in
\cite{Likhachev}.

Moreover, for the case $B_{\perp}= 100~kG$ the phase
in propagation factor reaches $\sim \pi$ for small mass parameters,
$\Delta m^2_5\sim 10^{-3}-10^{-5}$ resulting in zero ($\approx 0$)
of the conversion probability Eq. ({\ref{analytic}) while the survival
one remains at the level $P_{ee}\sim 1$. This is the reason why
right-handed $\bar{\nu}_{eR}$ are produced in a less amount for the
case $B_{\perp}=100~kG$ than in the case $B_{\perp} = 50~kG$.

This is also the reason why the
low mass region $\Delta m^2_5\sim 10^{-3}-10^{-5}$ which is allowed
from SK data for $B_{\perp}=50~kG$ (in Fig.\ \ref{fig10}) vanishes in the case
of $B_{\perp} = 100~kG$ (see Fig.\ \ref{fig11}).

For low magnetic fields (we built but do not present plots for regular
field $B_{\perp} = 20~kG$) only resonant $\nu_{eL}\to \bar{\nu}_{\mu
R}$-conversions take place in convective zone for $\Delta m^2\sim
10^{-7}~eV^2$ as well as MSW conversions for "large" mass parameters
$10^{-6}-10^{-5}~eV^2$. We can explain why this happens as follows.

These fields are not too large to suppress a resonance in
oscillation depth like for $B\sim 100~kG$.  On the other hand, for low
magnetic fields RSFP occurs as strong non-adiabatic conversion for
$\Delta m^2\sim 10^{-8}~eV^2$ in contrast to the adiabatic RSFP in the case
$B_{\perp} = 50~kG$ for the same mass parameter region.

Therefore, this
mass region is excluded from SK data in the case of low magnetic
fields (for $B= 20~kG$) and the triangle region $\Delta m^2\geq
10^{-7}~eV^2$ which is common for all cases is allowed only.

To resume the regular field case, we conclude that
parameter regions allowed from SK data are close to analogous ones obtained
in \cite{Akhmedov1} for small mixing angle band, $\sin^22\theta\lsim
0.25$.  However, in \cite{Akhmedov1} exclusion of large mixing angles
for small $\Delta m^2$ relevant to RSFP comes from fit with other
(radiochemical) experimental data  while in our case the corresponding
region in Fig. 10 (where $P_{\bar{e}\bar{e}}$ occurs to be noticeable) is
excluded from non-observation of antineutrinos $\bar{\nu}_{eR}$ in SK and
Kamiokande experiments \cite{Fiorentini} (see below).
\subsection{ The SK experiment bound on electron antineutrinos and
allowed parameter region}

In order to get antineutrino flux less than the
background in SK we should claim \cite{Fiorentini}
\widetext
\begin{equation}
\Phi_{\bar{\nu}_e}(E>8.3~MeV)
= \frac{\Phi_B^{SSM}(E>8.3)
\int_{8.3}^{15}\lambda_B(E)P_{\bar{e}\bar{e}}(E)
\sigma_{\bar{\nu}}(E)dE}{\int_{8.3}^{15}\lambda_B(E)\sigma_{\bar{\nu}}(E)dE}
 <~6\times 10^4~cm^{-2}s^{-1}~,
\label{antinu}
\end{equation}
\narrowtext

\noindent
where
$$\sigma_{\bar{\nu}}(E) = 9.2\times 10^{-42}~cm^2  [(E -
1.3~MeV)/10~MeV]^2$$
is the cross-section of the capture reaction
$\bar{\nu}_ep\to ne^+$;  $\Phi_B^{SSM}(E>8.3) =
\Phi_B^{(0)}\int_{8.3}^{15}\lambda_B(E)dE = 1.7\times
10^6~cm^{-2}s^{-1}$ is the SSM boron neutrino flux for chosen
threshold of sensitivity $E>E_0=8.3~MeV$\cite{Fiorentini}.

Deviding the inequality Eq. (\ref{antinu}) by this flux factor  we
find the bound on an averaged (over cross-section and spectrum)
transition probability $\nu_{eL}\to \bar{\nu}_{eR}$ (via a cascade),
\begin{equation}
\frac{\Phi_{\bar{\nu}_e}(E > 8.3~MeV)}{\Phi_B^{SSM}(E>8.3)}
= \frac{\int_{8.3}^{15}\lambda_B(E)P_{\bar{e}\bar{e}}(E)
\sigma_{\bar{\nu}}(E)dE}{\int_{8.3}^{15}\lambda_B(E)\sigma_{\bar{\nu}}(E)dE}
\lsim 0.035~,
\label{percent}
\end{equation}
or $\Phi_{\bar{\nu}}/\Phi_B^{SSM} \lsim 3.5$\%~.

We calculated boundaries in the parameter regions $\Delta m^2$,
$\sin^22\theta$ where the inequality Eq. (\ref{percent}) is violated in
order to exclude these regions from the allowed ones shown in
Figs.\ \ref{fig10}--\ref{fig14}.

We do not find any violation of the bound Eq. (\ref{percent}) for low
magnetic fields, $B~,b\lsim 20~kG$, both for regular and random ones.
Thus, we conclude that in the limit $B,~b\to 0$ (for instance for the MSW
case) the hole triangle region seen in Figs. 10-13 is allowed from the SK
data.

However, for strong magnetic fields such forbidden parameter regions
appear in {\it different} areas over $\Delta m^2$ and
$\sin^22\theta$ for {\it different} kind of magnetic fields ({\it
regular} and {\it random}) .  Moreover, for strong regular magnetic
fields ($B = 100~kG$) such dangerous parameter regions vanish (see
subsection IV.A above) while the stronger a random field would be the
wider forbidden area arises.

The bound Eq. (\ref{percent}) is not valid for low energy region below
the threshold of the antineutrino capture by protons, $E\leq 1.8~MeV$.

Therefore we can expect in future BOREXINO experiment a
large contribution of antineutrinos if neutrino has a large transition
magnetic moment $\mu\sim 10^{-11}\mu_B$. This is due to re-scaling of
the mass parameter $\Delta m^2$ that is not fixed and decrease of
neutrino energy that will be measured in BOREXINO. This prediction will
not be in contradiction with the bound Eq.(\ref{percent}) for hard
neutrinos\cite{PSV}.

Really, the higher the neutrino energy (or the smaller the
parameter $\delta = \Delta m^2/4E$) the less the probability $\langle
P_{\bar{e}\bar{e}}\rangle = \langle
\bar{\nu}_{eR}^*\bar{\nu}_{eR}\rangle$ occurs.
One can see from Figs.\ \ref{fig08}, \ref{fig09}
that for a wide region of the mixing angle
$\sin^22\theta$ and, for instance, for the neutrino mass parameter
$\Delta m^2\sim 10^{-8}~eV^2$ this probability is changing by 5-8
 times: having maximum $P_{\bar{e}\bar{e}}\sim 0.25-0.4$ at low
energies
 $E\lsim 1 MeV$ for $\sin^22\theta\gsim 0.06$ and the
minimum ($P_{\bar{e}\bar{e}}\lsim 0.05$)
at the energy region $E\gsim 5~MeV$.

\subsection{The MSW parameter region and correlation length of random
fields}

For "large" $\Delta m^2\sim 4\times 10^{-7}-10^{-5}~eV^2$ the MSW
conversions without helicity change, $\nu_{eL}\to \nu_{\mu L}$, take
place under the bottom of the convective zone as one can see from the
resonance condition Eq. (\ref{resonance}) where the neutron abundance
$Y_n$ (neutral current contribution) should be omitted.

Then left-handed neutrinos ($\nu_{eL}$~,$\nu_{\mu L}$) cross the
convective zone with random magnetic fields and can be converted to
right-handed components via the ASFC-mechanism. From Figs. 4, 7
it follows that the more intensive the r.m.s.  field
$\sqrt{\langle b^2\rangle}$ in the convective zone the more effective
spin-flavour conversions lead to the production of the right-handed
$\bar{\nu}_{eR}$, $\bar{\nu}_{\mu R}$-antineutrinos. This was also
proved analytically in subsection III.C
for small mixing angles both for $\delta$-correlated fields, Eq.
(\ref{(04)}), and for correlations of finite radius, Eq. (\ref{finite}).

Only "large" mass region $\Delta m^2\gsim 3\times
10^{-6}~eV^2-10^{-5}~eV^2$ survives as a pure MSW region without ASFC
influencing the left-handed components since our correlation length
$L_0=10^4~km$ chosen as a mesogranule size corresponds rather to finite
correlation radius (see subsection III.C.2).  Really, the dimensionless
parameter $x=\delta L_0$ is too large for these $\Delta m^2$ and the
ratio $\sin^2x/x^2$ ceases in Eq.  (\ref{finite}).  Therefore, there
are no ASFC for such mass parameters and $L_0=10^4~km$.

{\it Variation of the correlation length $L_0$}.
To check our approach we treated numerically a
more stronger r.m.s.  field $b=300~kG$ retaining invariant
$b^2L_0=const$ for a granule size $L_0\sim 10^3~km$, i.e. we considered
small-scale $\delta$-correlated random fields. In this case a lot of
$\bar{\nu}_{eR}$ appear even for the typical small-mixing MSW
region $\Delta m^2\sim 10^{-5}~eV^2$ excluding this and the whole
triangle region at all from non-observation of $\bar{\nu}_{eR}$ in SK.
This immediately follows from analytic formula Eq.  (\ref{finite})
since for $L_0=10^3~km$ limit ($\lim_{x\to 0}\sin x/x\to 1$) is fulfilled
there  and the auxiliary function $S$ ceases enhancing
$P_{{\nu}_{eL}\to \bar{\nu}_{\mu R}}$.

The same situation remains even for a more realistic
random magnetic field strength $b = 100~kG$  with $L_0\sim 10^3~km$,
see Fig.\ \ref{fig14} where we reconciled all solar neutrino experiments with
account of the bound  Eq. (\ref{percent})).

\section{Conclusions}

If antineutrinos $\bar{\nu}_{eR}$ would be found with the positive
signal in the Borexino experiment\cite{BOREXINO} or, in other words,
a small-mixing MSW solution to SNP fails, this will be a strong argument
in favour of magnetic field scenario with ASFC in the presence of a
large neutrino transition moment, $\mu\sim 10^{-11}\mu_B$ for the same
small mixing angle.

There appears one additional parameter for the
ASFC scenario comparing with the RSFP solution \cite{Akhmedov} to SNP.
This is the correlation length of random magnetic fields $L_0$ which
varies within the interval $L_0= 10^3-10^4~km$ in correspondence with typical
inhomogeneity size (of granules and mesogranules) in the Sun.

The probabilitues $P_{aa}$ sharply depend on the correlation length
$L_0$  that might allow (in the case of $\bar{\nu}_{eR}$ registered )
to study the structure of solar magnetic fields.

Another regular magnetic field parameter $\mu B_{\perp}$ \cite{Akhmedov}
changes  to $\mu \sqrt{\langle b^2\rangle}$ where
the r.m.s.  magnetic field $\sqrt{\langle b^2\rangle}$ was treated here
in the same interval $b= 20-100~kG$ as for usual estimates of regular
(toroidal) magnetic field in \cite{Akhmedov,Akhmedov1}.

Our main assumption about more stronger random field is based on modern
MHD models for solar magnetic fields where random fields are
naturally much bigger than large-scale magnetic fields created and
supported continuously from the small-scale random ones
\cite{Cataneo,Diamond} (see subsection  III.B).  The ratio
$b=\sqrt{R_m}B$\cite{Cataneo,Diamond} with large magnetic Reynolds
number $R_m\gg 1$ means that in RSFP scenarios (including twist
field model) the value of the regular large-scale field $B_{\perp}$ was
rather overestimated.

Thus, if neutrinos have a large transition magnetic moment
\cite{SchechterValle} their dynamics in the Sun is governed by random
magnetic fields that , first, lead to {\it aperiodic and rather
non-resonant} neutrino spin-flavor conversions, and second, inevitably
lead to production of electron antineutrinos for {\it low energy or
large mass difference} region.

The search of  bounds on $\mu$ at the level $\mu\sim 10^{-11}\mu_B$
in low energy $\nu~e$-scattering, currently planning in laboratory
experiments \cite{LAMA}, will be crucial for the model considered here.

We would like to emphasize the importance of future low-energy neutrino
experiments (BOREXINO, HELLAZ) which will be sensitive both to check
the MSW scenario and the $\bar{\nu}_{eR}$-production through ASFP.
As it was shown in a recent work\cite{PSV} a different slope of energy
spectrum profiles for different scenarios would be a crucial test in
favour of the very mechanism providing the solution to SNP.

\section*{ Acknowledgements}

The authors thank Sergio Pastor, Emilio Torrente, Jose Valle for
fruitful discussions. This work has been supported by RFBR grant
97-02-16501 and by INTAS grant 96-0659 of the European Union.


\vskip 2cm
\begin{figure}
\psfig{file=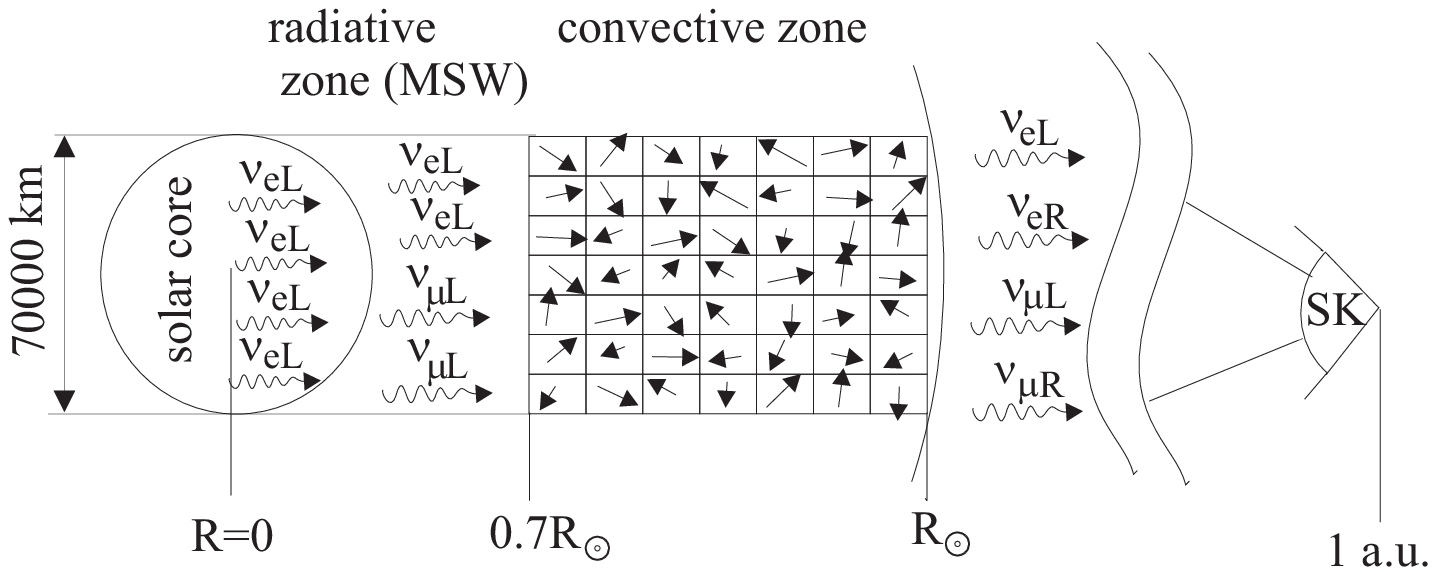}
\vskip0.8cm
\caption{Geometry of neutrino trajectories in random magnetic fields.
The finite radius of $\nu_{eL}$-neutrino source, the solar core radius
$0.1R_{\odot}$ is shown.}
\label{fig01}
\end{figure}

\newpage
\begin{figure}
\psfig{file=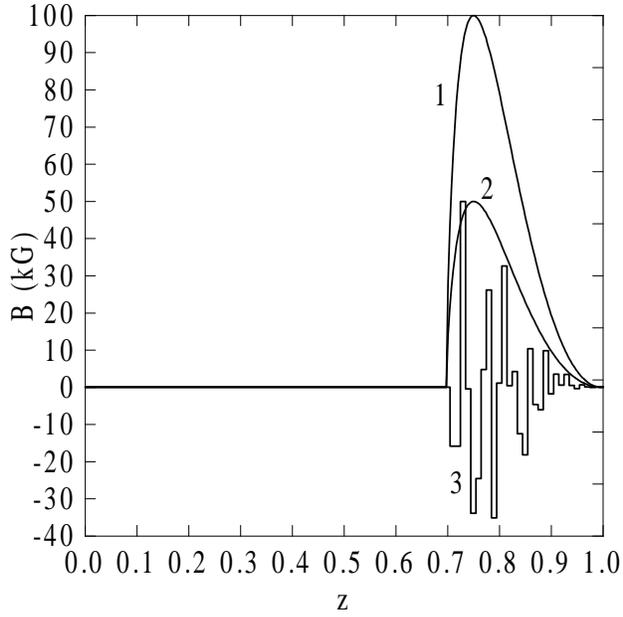,height=8cm,width=8cm}
\vskip0.8cm
\caption{Profiles of regular and random magnetic fields in the solar
convective zone. One realization of random magnetic field component
$b_x(r)$ (for one neutrino ray) is plotted. $b_y(r)$ component which
has a different profile is not shown.
1~-~regular magnetic field profile with $B_{max} = 100~kG$,
2~-~the same with $B_{max} = 50~kG$,
3~-~random magnetic field profile with $b_{rms} = 50~kG$. }
\label{fig02}
\end{figure}
\newpage
\begin{figure}
\psfig{file=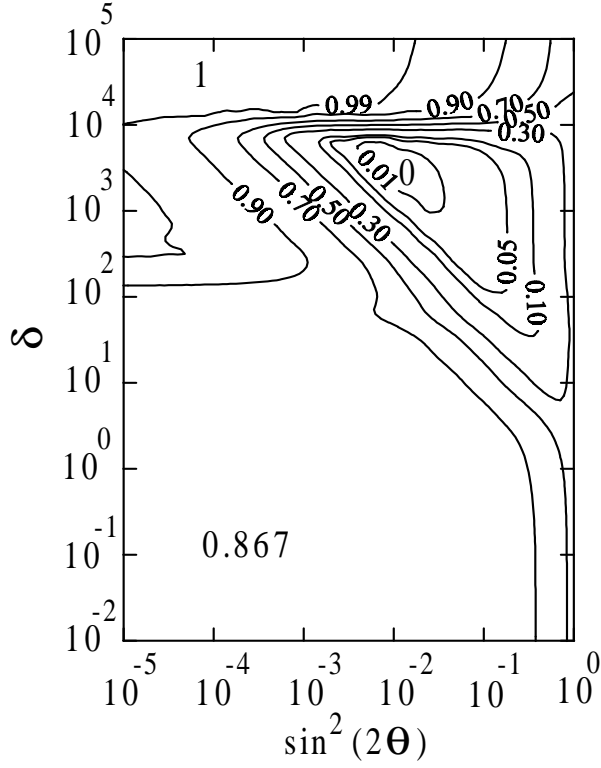,height=10.18cm,width=8cm}
\vskip0.8cm
\caption{The averaged (over 50 rays) survival probability
$P_{ee} = \langle\nu_{eL}^*\nu_{eL}\rangle$ in dependence on
$\delta = \Delta m^2/4E = 8.81\times10^8~\Delta m^2 (eV^2)/E(MeV)$
and $\sin^22\theta$ -- parameters
for $b_{r.m.s.}= 50~kG$ and correlation length $L_0 = 10^4~km$.}
\label{fig03}
\end{figure}
\newpage
\begin{figure}
\psfig{file=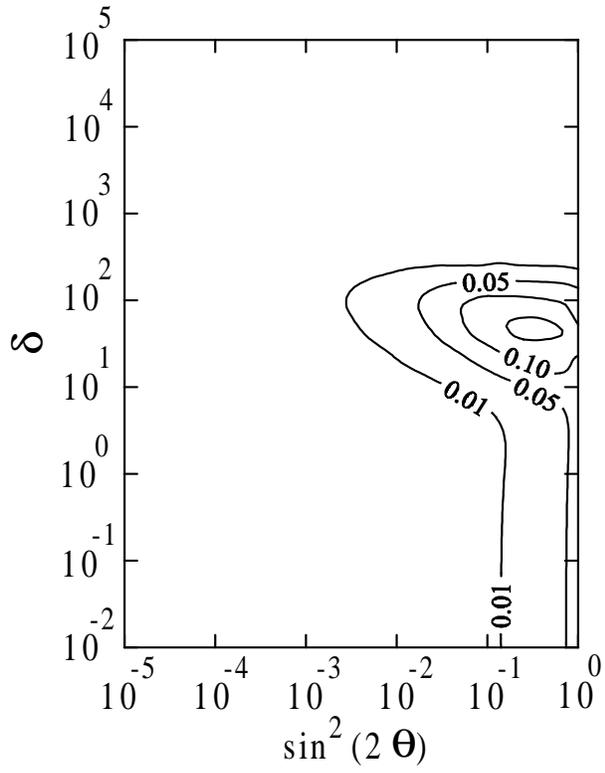,height=10.18cm,width=8cm}
\vskip0.8cm
\caption{As previous figure. The average probability
$P_{\bar{e}\bar{e}} = \langle \bar{\nu}_{eR}^*\bar{\nu}_{eR}\rangle$.}
\label{fig04}
\end{figure}
\newpage
\begin{figure}
\psfig{file=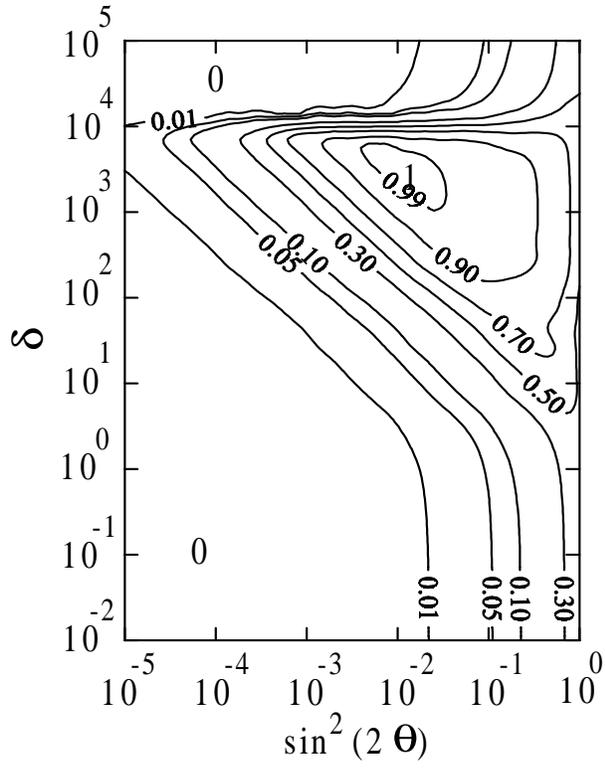,height=10.18cm,width=8cm}
\vskip0.8cm
\caption{As previous figure. The average probability
$P_{\mu\mu} = \langle \nu_{\mu L}^*\nu_{\mu L}\rangle$.}
\label{fig05}
\end{figure}
\newpage
\begin{figure}
\psfig{file=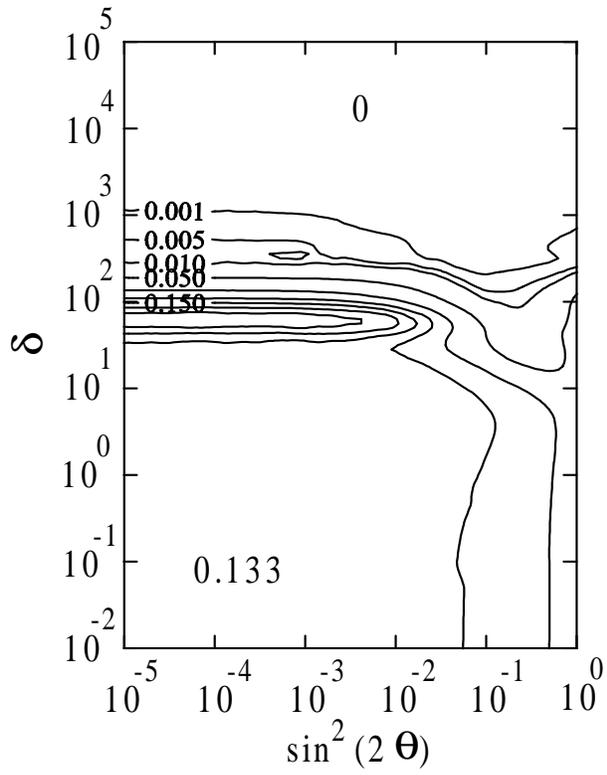,height=10.18cm,width=8cm}
\vskip0.8cm
\caption{As previous figure. The average probability
$P_{\bar{\mu}\bar{\mu}} = \langle
\bar{\nu}_{\mu R}^*\bar{\nu}_{\mu R}\rangle$.}
\label{fig06}
\end{figure}
\newpage
\begin{figure}
\psfig{file=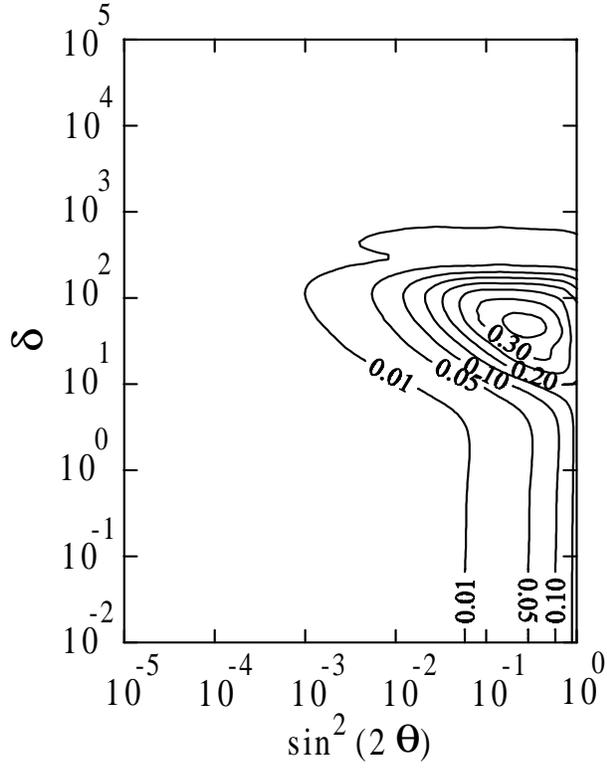,height=10.18cm,width=8cm}
\vskip0.8cm
\caption{As previous figure. The average probability
$P_{\bar{e}\bar{e}} = \langle \bar{\nu}_{eR}^*\bar{\nu}_{eR}\rangle$
for $b_{r.m.s.}= 100~kG$ and correlation length $L_0 = 10^4~km $.}
\label{fig07}
\end{figure}
\newpage
\begin{figure}
\psfig{file=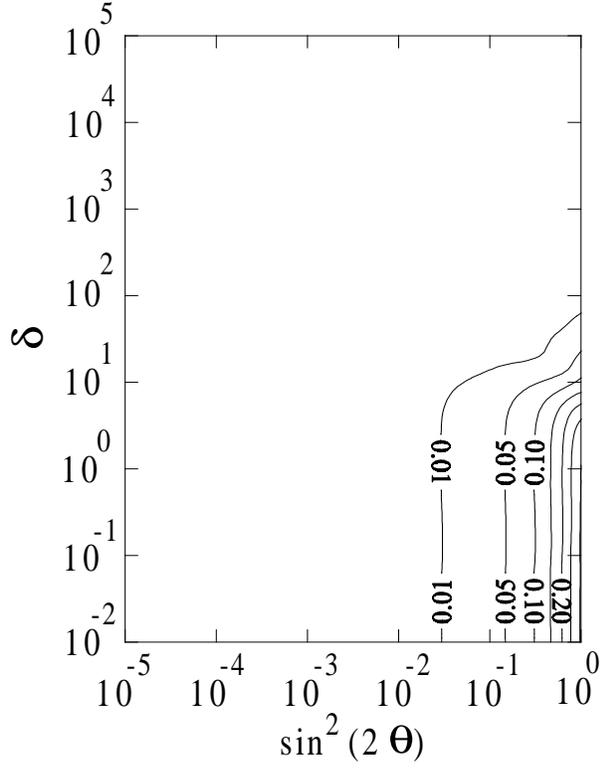,height=10.18cm,width=8cm}
\vskip0.8cm
\caption{The  probability $P_{\bar{e}\bar{e}} =
\langle \bar{\nu}_{eR}^*\bar{\nu}_{eR}\rangle$ in dependence on $\delta
= \Delta m^2/4E$ and $\sin^22\theta$-parameters for
the regular field $B_0= 50~kG$.}
\label{fig08}
\end{figure}
\newpage
\begin{figure}
\psfig{file=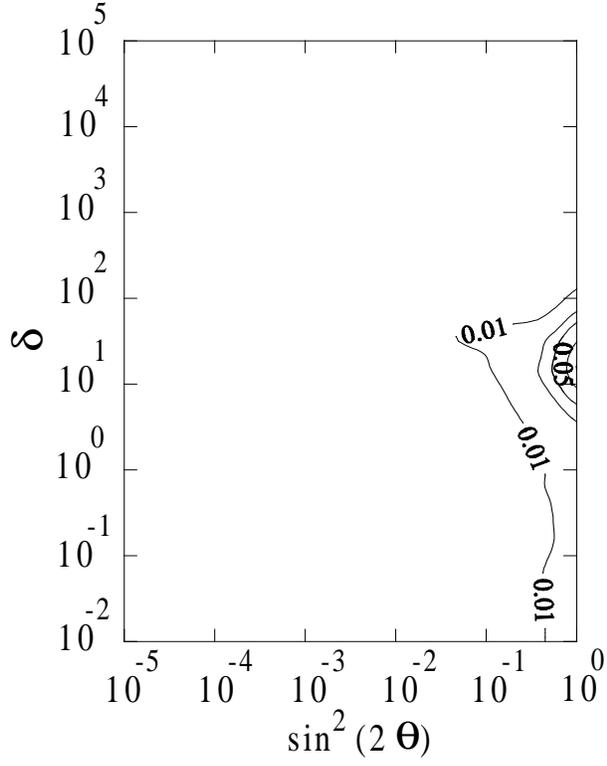,height=10.18cm,width=8cm}
\vskip0.8cm
\caption{As previous figure. The average probability
$P_{\bar{e}\bar{e}} = \langle \bar{\nu}_{eR}^*\bar{\nu}_{eR}\rangle$
for the regular field $B_0= 100~kG$.}
\label{fig09}
\end{figure}
\newpage
\begin{figure}
\psfig{file=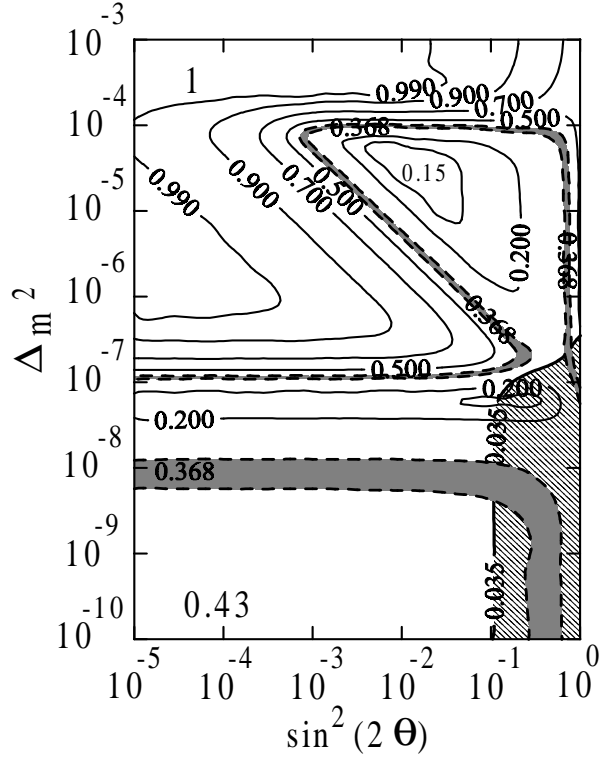,height=10.18cm,width=8cm}
\vskip0.8cm
\caption{$\Delta m^2~(eV^2)$, $\sin^22\theta$ -plane for the ratio
DATA/SSM in the regular field $B_0= 50~kG$. The isoline $DATA/SSM =
0.368\pm 0.026$ allowed from the SK experiment is shown by the dashed
line.  The SK bound on $\bar{\nu}_{eR}$ Eq.  (\ref{percent}) is
plotted too separating the excluded area.}
\label{fig10}
\end{figure}
\newpage
\begin{figure}
\psfig{file=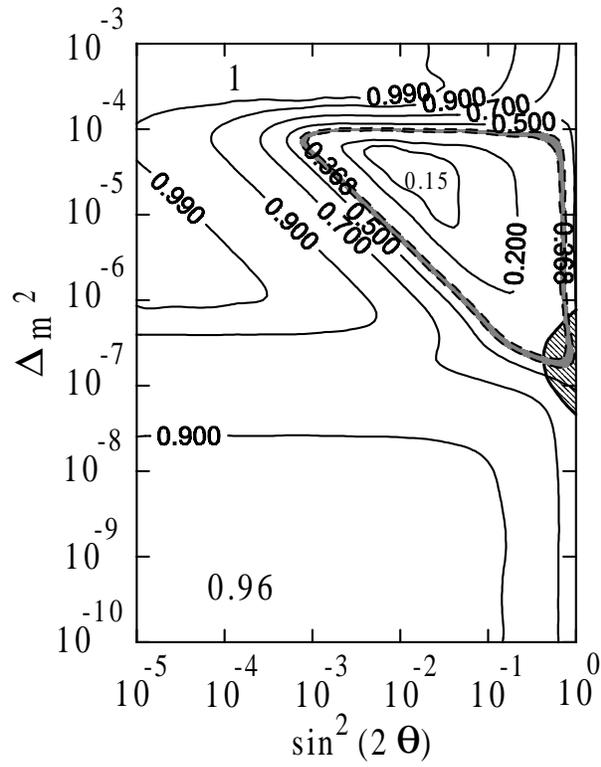,height=10.18cm,width=8cm}
\vskip0.8cm
\caption{As previous figure for the ratio DATA/SSM in the
regular field $B_0= 100~kG$.}
\label{fig11}
\end{figure}
\newpage
\begin{figure}
\psfig{file=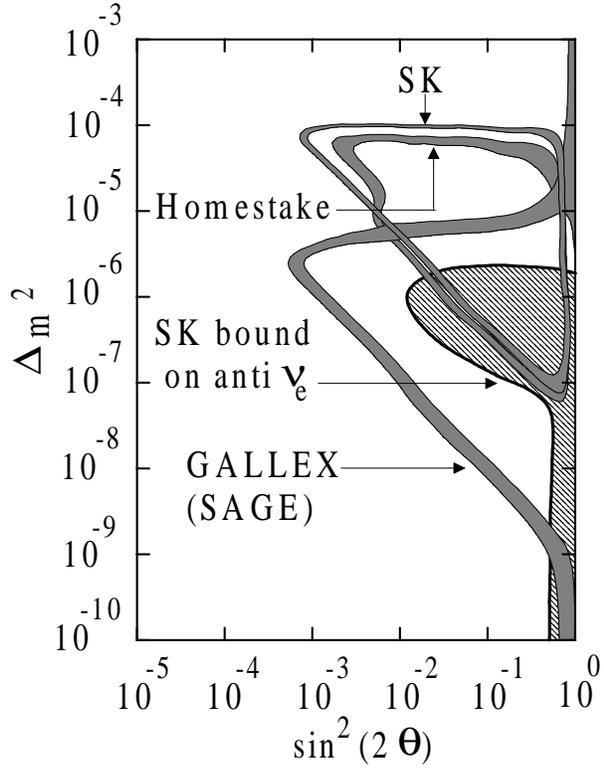,height=10.18cm,width=8cm}
\vskip0.8cm
\caption{The reconcilement of all experiments. The isolines with
dashed bands represent  the DATA/SSM ratio equal correspondingly to
$0.368\pm 0.026~$(SK),   $0.509\pm 0.059$,  $0.504 \pm 0.085$
(GALLEX+SAGE),  and $0.274\pm 0.027~$(Homestake).
The random field $b_{rms}= 50~kG$ and correlation length $L_0 = 10^4~km $.}
\label{fig12}
\end{figure}
\newpage
\begin{figure}
\psfig{file=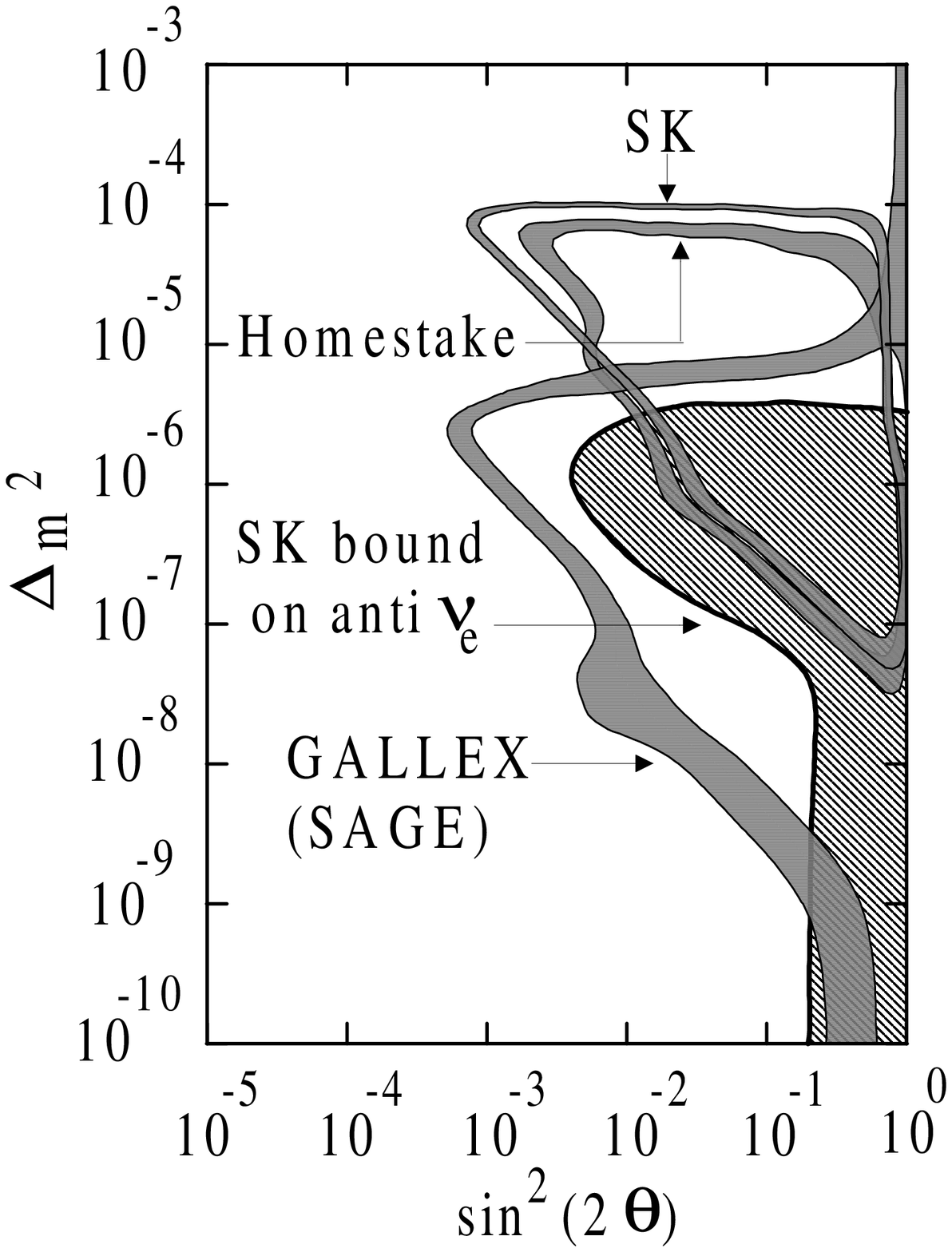,height=10.18cm,width=8cm}
\vskip0.8cm
\caption{As previous figure.
The random field $b_{rms}= 100~kG$ and correlation length $L_0 = 10^4~km $.}
\label{fig13}
\end{figure}
\newpage
\begin{figure}
\psfig{file=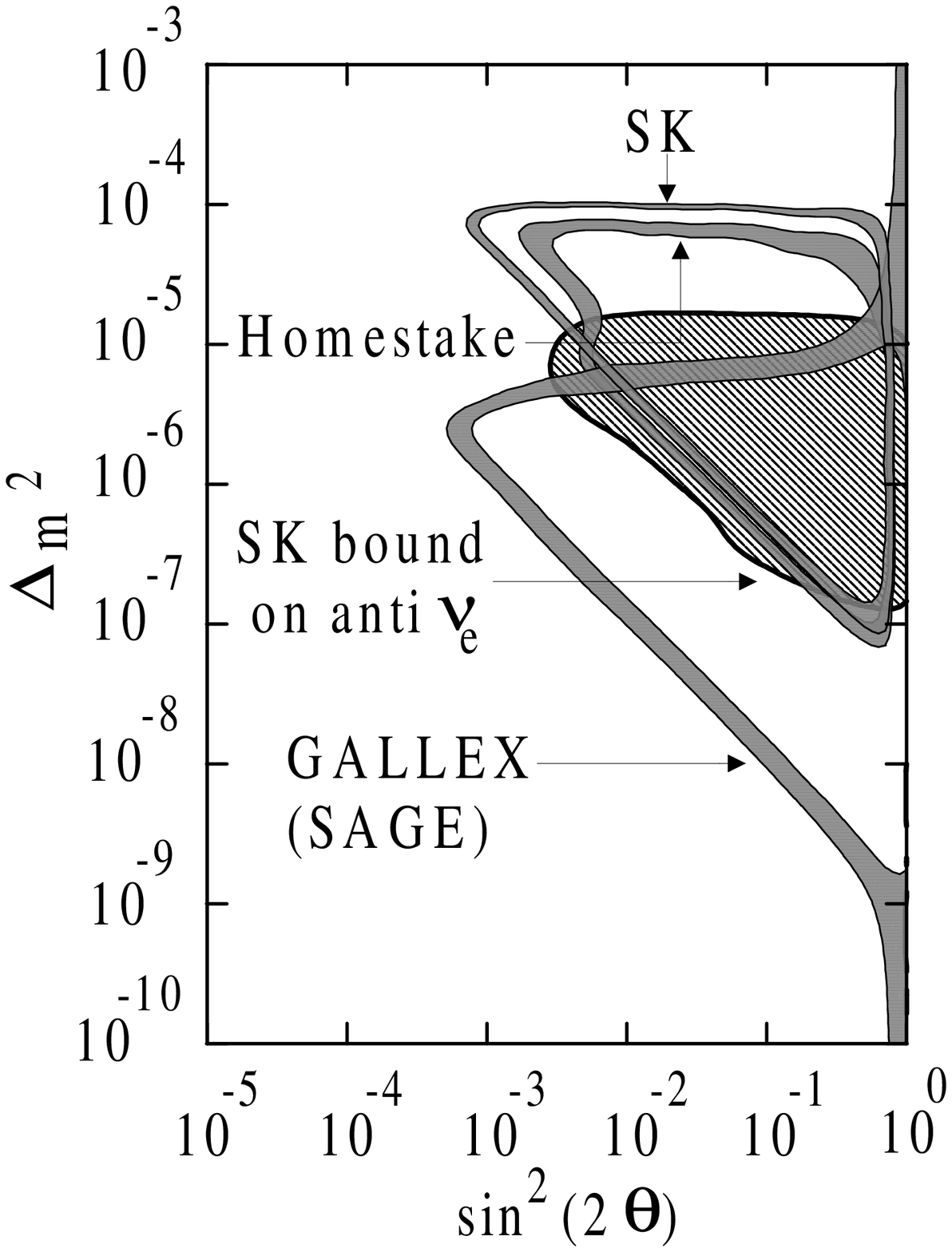,height=10.18cm,width=8cm}
\vskip0.8cm
\caption{As previous figure.
The random field $b_{rms}= 100~kG$ and correlation length $L_0 = 10^3~km $.}
\label{fig14}
\end{figure}


\begin{thebibliography}{99}
\bibitem{SK}

Y. Suzuki, Rapporter talk at 25th ICRC (Durban) 1997.

\bibitem{Berezinsky}
V. Berezinsky, astro-ph/9710126, invited lecture at
25th International Cosmic Ray Conference, Durban,
28 July - 8 August, 1997.

\bibitem{Parker}
E.N. Parker, Astrophys. J., 408 (1993) 707;\\
E.N. Parker, {\em Cosmological Magnetic Fields},
Oxford University Press, Oxford, 1979.

\bibitem{Stix}
S.I. Vainstein, A.M. Bykov, I.M. Toptygin, {\em Turbulence,
Current Sheets and Shocks in Cosmic Plasma}, Gordon and Breach, 1993.

\bibitem{Nicolaidis}
A.Nicolaidis,Phys.Lett., B262 (1991) 303.

\bibitem{Enqvist}
K. Enqvist, P. Olesen, V.B. Semikoz, Phys. Rev. Lett. 69 (1992) 2157\\
K.Enqvist, A.I. Rez, V.B. Semikoz, Nucl. Phys. B436 (1995) 49;\\
S. Pastor, V.B. Semikoz, J.W.F. Valle, Phys. Lett. B369 (1996) 301.

\bibitem{Akhmedov}
E. Kh. Akhmedov, Phys. Lett. B 213 (1988) 64;\\
C.-S. Lim and W.J. Marciano, Phys. Rev. D 37 (1988) 1368.

\bibitem{Akhmedov1}
E.Kh. Akhmedov, ``The neutrino magnetic moment and time variations
of the solar neutrino flux'', Preprint IC/97/49, Invited talk given
at the 4-th International Solar Neutrino Conference, Heidelberg,
Germany, April 8-11, 1997.

\bibitem{Fiorentini}
G. Fiorentini, M. Moretti and F.L. Villante, hep-ph/9707097

\bibitem{APS}
E.Kh. Akhmedov, S.T. Petcov and A.Yu. Smirnov, Phys. Rev. D48 (1993)
2167; Phys. Lett. B309 (1993) 95.

\bibitem{PSV}
S. Pastor, V.B. Semikoz and J.W.F. Valle, Phys. Lett. B423 (1998) 118;
hep-ph/9711316

\bibitem{BL}
A.B. Balantekin and F. Loreti, Phys. Rev. D48 (1993) 5496.

\bibitem{Bahcall}
John N. Bahcall, {\em Neutrino Astrophysics}, Cambridge University
Press, 1988, section 6.3.

\bibitem{Lifshitz}
I.M.Lifshitz, S.A.Gredeskul, and L.A.Pastur, Sov. Phys. JETF 83 (1982) 2362
(in Russian).

\bibitem{Sem97}
V.B. Semikoz, Nucl.Phys. B501 (1997) 17

\bibitem{radiochem}
J.K. Rowley et al. {\em in Solar Neutrinos and Neutrino Astronomy},
AIP Conference Proceedings, 126, edited by M.L. Cherry, W.A. Fowler
and K. Lande, (1985);\\\
P.Anselman et al. Phys. Lett. B327 (1994) 377;\\
J.N.Abdurashitov et al. Phys.Lett. B328 (1994) 234.

\bibitem{BOREXINO}
Status Report of Borexino Project: {\it The Counting Test Facility
and
its Results. A proposal for participation in the Borexino Solar
Neutrino Experiment}, J.B. Benziger et al., (Princeton, October
1996).
Talk presented by P.A. Eisenstein at Baksan International School
{\it"
Particles and Cosmology"} (April 1997).

\bibitem{HELLAZ} F. Arzarello et al. Preprint CERN-LAA/94-19,
College de France LPC/94-28; \\ C. Laurenti et al. {\it Proceedings
of
the 5th Int. Workshop on Neutrino Telescopes, Venice XXX};\\ G.
Bonvicini, Nucl.  Phys. B (Proc. Suppl.) 35 (1994) 438.

\bibitem{BP}
J.N. Bahcall, M.H. Pinsonneault, Rev. Mod. Phys. 67 (1995) 781.

\bibitem{Grimus}
W.Grimus and T. Scharnagl, Mod. Phys. Lett. A8 (1993) 1943

\bibitem{Zeldovich}
Ya.A. Zeldovich, A.A. Ruzmaikin, D.D. Sokoloff, {\em Magnetic
fields in astrophysics}, Cordon and Breach, N.Y., 1983

\bibitem{Ruzmaikin}
A.A. Ruzmaikin, A.M. Shukurov, D.D. Sokoloff, {\em Magnetic fields
of Galaxies}, Kluwer, Dordrecht, 1988.

\bibitem{Cataneo}
S.I. Vainstein, F. Cattaneo, 1992, Astrophysical Journal, 393 (1992)
165.

\bibitem{Diamond}
A.V. Gruzinov, P.H. Diamond, 1994, Phys. Rev. Lett.,  72 (1994) 1651.

\bibitem{Abrikosov}
A.A. Abrikosov, L.P. Gorkov, I.E. Dzyaloshinski,
{\em Methods of Quantum Field in Statistical Physics}, Prentice Hall, 1963.

\bibitem{Balantekin}
A.B.Balantekin, J.M.Fetter, and F.N.Loreti, Phys. Rev. D 54 (1996) 3941.

\bibitem{Likhachev}
G.G. Likhachev, A.I. Studenikin, Sov. Phys. JETP, 81 (1995) 419.

\bibitem{Krastev}
J.N. Bahcall, P. Krastev, Preprint IASSNS-AST 97/31.

\bibitem{SchechterValle}
J. Schechter, J.W.F. Valle, Phys. Rev. D 24 (1981) 1883;
Err. Phys. Rev. D 25 (1982) 283.

\bibitem{LAMA}
I.R. Barabanov et al., Astroparticle Phys., 5 (1996) 159.
\end{thebibliography}
\end{document}